\documentclass[sigconf,screen]{acmart}
\AtBeginDocument{%
  }

\setcopyright{none}

\acmConference[Preprint]{}{2026}{}
\acmISBN{}                       
\acmDOI{}                        
\acmYear{}                       
\acmPrice{}

\acmSubmissionID{1987} 

\author{Ye Yu}
\affiliation{%
  \institution{Columbia University}
  \city{New York}
  \state{NY}
  \country{USA}
}
\email{yy3628@columbia.edu}

\author{Yifan Zhou}
\affiliation{%
  \institution{Peking University}
  \city{Beijing}
  \state{Beijing}
  \country{China}
}
\email{welementzyf@stu.pku.edu.cn}

\author{Yi Chen}
\affiliation{%
  \institution{Peking University}
  \city{Beijing}
  \state{Beijing}
  \country{China}
}
\email{yichen25@stu.pku.edu.cn}

\author{Pedro Soto}
\affiliation{%
  \institution{Virginia Tech}
  \city{Blacksburg}
  \state{Virginia}
  \country{USA}
}
\email{pedro.juan.soto.conde@gmail.com}

\author{Wenjie Xiong}
\affiliation{%
  \institution{Virginia Tech}
  \city{Blacksburg}
  \state{Virginia}
  \country{USA}
}
\email{wenjiex@vt.edu}

\author{Meng Li}
\affiliation{%
  \institution{Peking University}
  \city{Beijing}
  \state{Beijing}
  \country{China}
}
\email{meng.li@pku.edu.cn}

\definecolor{darkblue}{RGB}{0,10,150}
\newcommand{\supp}[1]{{#1}}

\newcommand{\method}{Cachemir}

\usepackage{tikz}
\usepackage[normalem]{ulem}

\usepackage{amsmath,amssymb,amsfonts}
\usepackage{algorithmic}
\usepackage{graphicx}
\usepackage{textcomp}
\usepackage{xcolor}

\usepackage{url}            
\usepackage{algorithm2e}
\RestyleAlgo{ruled}

\usepackage{adjustbox}
\usepackage{makecell}
\usepackage{booktabs}
\usepackage{multirow}
\usepackage{bbm}
\usepackage{subfig}
\usepackage[flushleft]{threeparttable}
\usepackage{pifont}
\begin{document}

\title{\method: Fully Homomorphic Encrypted Inference of Generative Large Language Model with KV Cache}



\begin{abstract}



Generative large language models (LLMs) have revolutionized multiple domains. Modern LLMs predominantly rely on an autoregressive decoding strategy, which generates output tokens sequentially and employs a key-value cache (KV cache) to avoid redundant computation. However, the widespread deployment of LLMs has raised serious privacy concerns, as users are feeding all types of data into the model, motivating the development of secure inference frameworks based on fully homomorphic encryption (FHE). A major limitation of existing FHE-based frameworks is their inability to effectively integrate the KV cache, resulting in prohibitively high latency for autoregressive decoding.
In this paper, we propose \method, a KV \underline{C}ache \underline{AC}celerated \underline{H}omomorphic \underline{E}ncrypted LL\underline{M} \underline{I}nference \underline{R}egime to overcome this limitation. \method~comprises three key technical contributions: 1) a set of novel HE packing algorithms specifically designed to leverage the computational advantages of the KV cache; 2) an interleaved replicated packing algorithm to efficiently compute the vector-matrix multiplications that result from using the KV cache in Transformer linear layers; and 3) an augmented bootstrapping placement strategy that accounts for the KV cache to minimize bootstrapping cost. We demonstrate that \method~achieves $48.83\times$ and $67.16\times$ speedup over MOAI (ICML'25) and THOR (CCS'25) respectively on CPU and consumes less than 100 seconds on GPU to generate an output token for Llama-3-8B. Our code is available at \hyperlink{here}{https://anonymous.4open.science/r/Cachemir-B254}.

\end{abstract}





\maketitle

\section{Introduction}
\label{sec:intro}

Generative large language models (LLMs), e.g., GPT \cite{chatgpt}, LLAMA \cite{touvron2023llama}, etc, have demonstrated remarkable capability in a wide range of applications, including question answering, document summarization and editing \cite{koh2022summarize}, financial systems \cite{ding2024llmtrading}, medical diagnosis \cite{shamshad2022transformersmedicalimagingsurvey}, etc. As existing LLM services are mostly deployed on the cloud, users are required to upload their prompts, which may involve sensitive personal information. Therefore, privacy has become a major concern for many \cite{huang2022cheetah,mishra2020delphi,rathee2021sirnn,rathee2020cryptflow2}.

\begin{figure}[!tb]
    \centering
    \includegraphics[width=\linewidth]{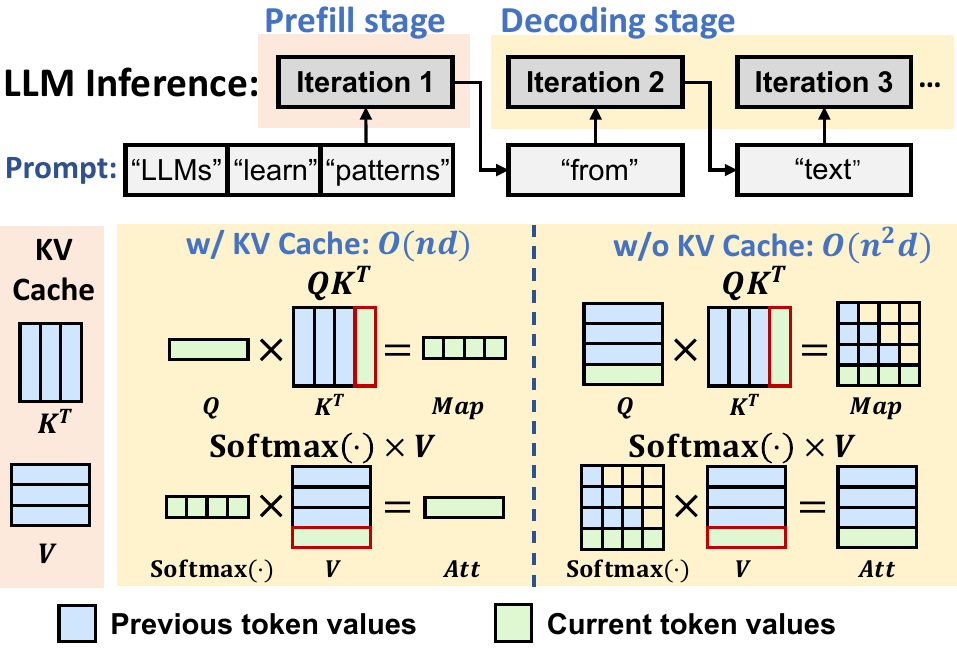}
    \caption{Auto-regressive decoding for generative inference of LLMs. The computation of $QK^T$ and $\mathrm{Softmax}(\cdot)\times V$ for previous tokens (in blue) can be saved with KV cache. 
    }
    \label{fig:intro}
\end{figure}

To address these privacy concerns, secure inference frameworks have attracted considerable research interest. Existing approaches can be broadly categorized into three paradigms: fully homomorphic encryption (FHE) based \cite{Dowlin2016CryptoNetsAN,chen-etal-2022-thex,Kim2023convfhe,Ebel2023OrionAF,park2024powerformer,moon2024thor,zhang2025moai}, secure multi-party computation (MPC) based \cite{Zeng2022MPCViTSF,dong2023puma,akimoto2023privformer,li2022mpcformer,gupta2023sigma,keic25shaft}, and hybrid frameworks combining HE and MPC \cite{juvekar2018gazelle,mishra2020delphi,huang2022cheetah,hao2022iron,pang2023bolt,lu2023bumblebee,He2024RhombusFH,ju2024neujeans,hou2023ciphergpt,xu2024privcirnet}. Both MPC-based and hybrid frameworks suffer from several drawbacks. First, the frameworks either incur significant communication overhead due to their reliance on cryptographic primitives like oblivious transfer (OT), or require a non-colluding third party. Moreover, they introduce non-negligible computation load on the client. 
In contrast, an FHE-based framework is much more communication-efficient and does not require assumptions on non-colluding parties: after the initial exchange of encrypted inputs, all computation occurs on the untrusted server-side before the final encrypted result is returned to the client. Therefore, in this paper, FHE-based secure LLM inference becomes our primary focus.

Generative LLM inference typically consists of two distinct stages: 1) the prefill stage, which processes the entire input prompt concurrently, and 2) the decoding stage, which generates output tokens sequentially in an autoregressive manner \cite{Radford2018gpt}. As shown in Figure~\ref{fig:intro}, a core operation in LLMs is self-attention, calculated as $\mathrm{Softmax}(QK^\top)V$, where $Q, K, V$ denote the query, key, and value matrices, respectively (detailed in Section~\ref{subsec:back:llm}). In the decoding stage, the query vector of each new token must attend to the key and value vectors of all previous tokens. To avoid recomputation, a key-value cache (KV cache) is commonly employed to store all these previous states. Therefore, each decoding step only involves processing a single input token, reducing the main computation from a matrix multiplication to a vector-matrix multiplication (VMM).


However, prior-art FHE-based secure inference frameworks, e.g., NEXUS \cite{Zhang2024SecureTI} and THOR \cite{moon2024thor}, primarily focus on non-autoregressive language models like BERT \cite{Devlin2019BERTPO} or the prefill stage of autoregressive LLMs \cite{Radford2018gpt}. When directly adopting these frameworks to support LLM decoding, we observe emerging challenges as discussed below:

\ding{182} \textbf{Inefficient VMM Protocol.} As LLMs compute over tensors while FHE operates on ciphertext polynomials, efficient packing, which encodes tensors into one-dimensional polynomial coefficients, is important for both correctness and performance of encrypted computation and has been studied by prior works \cite{juvekar2018gazelle,huang2022cheetah}. 

Existing FHE frameworks have primarily optimized packing for convolutions \cite{Kim2023convfhe,Ebel2023OrionAF} and matrix-matrix multiplications \cite{Zhang2024SecureTI,moon2024thor,park2024powerformer}, but lack efficient support for VMMs prevalent in LLM decoding. The gap stems from a fundamental mismatch: FHE polynomial degrees (typically $2^{15}$ and $2^{16}$ \cite{Zhang2024SecureTI,zhang2025moai}) far exceed the hidden dimensions of common LLMs ($2^{10}$ to $2^{13}$ \cite{grattafiori2024llama3herdmodels,radford2019languagegpt2,touvron2023llama}). A naive solution is to pad tensors with zeros to fit the ciphertext size, which, however, not only wastes substantial computation on padded ciphertext slots, but also introduces extra operations, e.g., homomorphic rotations, to maintain computation correctness \cite{peng2023lingcn}. An alternative, proposed by MOAI \cite{zhang2025moai}, packs multiple data batches into a single ciphertext. Yet, this strategy struggles to handle the variable sequence lengths inherent to the prefill and decoding phases. Moreover, it also increases the memory consumption to store the model proportionally to the batch size. Consequently, designing efficient packing algorithms for VMMs remains a significant challenge.


\ding{183} \textbf{Incompatible with KV Cache Management.} Although the KV cache drastically reduces computational overhead in autoregressive decoding, its dynamic nature—growing cache size with each decoding step—renders prior FHE packing algorithms incompatible. For instance, diagonal packing in THOR \cite{moon2024thor} and interleaved batching in MOAI \cite{zhang2025moai} assume a static KV cache size and thus cannot accommodate its incremental updates. Similarly, the offline-online matrix multiplication in NEXUS \cite{Zhang2024SecureTI} is designed for fixed-size weight matrices, not the evolving state of the KV cache. Hence, enabling efficient autoregressive decoding requires new cryptographic protocols that natively support dynamic KV cache updates and the subsequent computations.


\ding{184} \textbf{Suboptimal Bootstrapping Placement.} As each FHE ciphertext supports only a finite number of operations before accumulating excessive noise, bootstrapping is required to reset its noise level and enable further computation \cite{cheon2018bootstrapping}. The placement of bootstrapping operations is critical, as it is computationally intensive, directly impacts the computational cost of subsequent FHE operations and constrains future bootstrapping choices.

Although prior work has explored both manual and automatic bootstrapping placement strategies \cite{Zhang2024SecureTI, moon2024thor, zhang2025moai, Ebel2023OrionAF, compiler_dacapo, compiler_fhelipe, compiler_helayers}, applying them directly to the KV-cache-enabled decoding stage is challenging and suboptimal. First, existing algorithms (e.g., in NEXUS \cite{Zhang2024SecureTI} and Orion \cite{Ebel2023OrionAF}) restrict placement to the boundaries between network layers\footnote{A network layer is an operation such as a linear transformation (e.g., convolution, matrix multiplication) or a nonlinear function (e.g., GeLU, LayerNorm) \cite{Ebel2023OrionAF}.}, as the prefill stage produces numerous intermediate ciphertexts in nonlinear layers. In contrast, the decoding stage with a KV cache allows the intermediate tensors of a nonlinear layer to be packed into a single ciphertext, creating new opportunities to bootstrap within a layer and expanding the solution space drastically for placement. Second, the deeper and more complex architecture of modern LLMs further complicates the placement problem. Prior approaches, such as the shortest-path algorithm used in Orion \cite{Ebel2023OrionAF}, struggle with high algorithmic complexity and prohibitive runtime when applied to these large graphs.

\subsection{Technical Details}

To overcome the limitations of existing solutions, we propose \method, a KV \underline{C}ache \underline{AC}celerated \underline{H}omomorphic \underline{E}ncrypted LL\underline{M} \underline{I}nference \underline{R}egime, which features three key innovations to resolve the challenges above:

\ding{182} \textbf{Efficient VMM Packing.} To address the severe ciphertext underutilization for VMMs, we introduce a novel \textbf{Interleaved Replicated Packing} algorithm. Its key innovation is to replicate input vectors and pack them into a single ciphertext in an interleaved pattern. This layout maximizes ciphertext slot utilization, which simultaneously reduces the number of expensive homomorphic rotations and decreases multiplicative depth consumption. Although the valid results must be extracted from the output ciphertexts to ensure the correctness for subsequent layers, we demonstrate that this extraction can be fused with operations in subsequent layers. Consequently, our packing algorithm reduces the multiplicative depth of a VMM to a single level, while reducing the number of rotations by $2521\times$ and $7.2\times$ compared to NEXUS \cite{Zhang2024SecureTI} and an adapted version of BOLT \cite{pang2023bolt}\footnote{While BOLT only uses small polynomial degree ($2^{13}$), we adapt their method via zero-padding to support larger degrees ($2^{16}$).}, respectively.

\ding{183} \textbf{Novel Protocol for KV Cache Computation.} We propose the first efficient protocol to support homomorphic update and computation of KV cache during autoregressive decoding. Specialized packing schemes are designed for the \textbf{K Cache} and \textbf{V Cache}, along with novel protocols for their respective ciphertext multiplications, ensuring full compatibility with our Interleaved Replicated Packing.

For the \textbf{K Cache}, key vectors from multiple tokens are packed into a single ciphertext following an interleaved layout. This minimizes storage and enables parallel computation of the $QK^\top$ dot product (illustrated in Figure~\ref{fig:intro}). For the \textbf{V Cache}, we design a complementary strategy where each value vector is distributed across multiple ciphertexts. This layout allows the subsequent $\mathrm{Softmax}(\cdot)\times V$ operation (illustrated in Figure~\ref{fig:intro}) to natively produce outputs in the interleaved format required by downstream layers. Moreover, they introduce negligible updating overhead.  

Critically, our packing algorithms support appending new key and value vectors to the encrypted KV cache with \textbf{negligible computational overhead}, thereby natively supporting autoregressive decoding with variable sequence lengths. Compared to the state-of-the-art (SOTA) method of THOR \cite{moon2024thor}, which does not use KV cache, this method can reduces the inference runtime by $11\sim133\times$.

\ding{184} \textbf{Efficient bootstrapping placement.} We also propose a new bootstrapping placement algorithm that enhances the approach of Orion \cite{Ebel2023OrionAF}, overcoming its suboptimality and high complexity for LLMs. Our method is driven by two key observations. Firstly, the number of ciphertexts remains almost consistent across linear and nonlinear layers within the decoding stage. This allows us to break traditional layer boundaries and insert bootstrapping operations within a single nonlinear layer, significantly increasing placement flexibility. Second, LLMs are composed of repeated Transformer blocks with identical dimensions. To manage the expanded solution space without prohibitive complexity, we prune the search space by optimizing placement within a single representative Transformer block and then replicating this strategy across all blocks. This approach reduces bootstrapping latency by $1.98\times$ for Llama-3-8B. 


\subsection{Contributions}
Our contributions are summarized as follows:
\begin{enumerate}
    \item  We propose \textbf{Interleaved Replicated Packing}, a novel packing scheme for VMM in the decoding stage. It reduces the number of homomorphic rotations by $7.2\sim 2521\times$ compared to SOTA methods, including BOLT\cite{pang2023bolt}, MOAI\cite{zhang2025moai}, THOR\cite{moon2024thor}, and NEXUS\cite{Zhang2024SecureTI}. 
    \item We propose the \textbf{first} FHE protocol for dynamic KV cache management that natively supports variable sequence lengths in autoregressive decoding, enabling practical and efficient encrypted LLM inference.
    \item We introduce a new bootstrapping placement algorithm tailored for KV-cache-enabled decoding. By enabling bootstrapping within nonlinear layers and solution space pruning, our method achieves 1.98$\times$ latency reduction over SOTA algorithm Orion\cite{Ebel2023OrionAF}.
    \item We integrate these techniques into an FHE-based inference framework, \method. Extensive results demonstrate that \method~achieves a latency of less than $100$ seconds per generated token on the Llama-3-8B model with GPU acceleration and outperforms MOAI and THOR by $48.83\times$ and $67.16\times$ on CPU, respectively.
    
\end{enumerate}

\section{Background}
\label{sec:back}

In this section, we provide the necessary background of
this paper. Table \ref{tab:notation} shows the frequently used notations.

\begin{table}[!tb]
\caption{Frequently used notations}
\label{tab:notation}
  \footnotesize
  \begin{threeparttable}
    \begin{tabular}{@{}c@{}c|c@{}c@{}}
    \toprule
    \textbf{} & \textbf{Description} & \textbf{} & \textbf{Description} \\
    \midrule
    $\mathbf{p}$ &  plaintext & $\mathbf{c}$ &  ciphertext \\
    $\oplus$ & homomorphic addition & $\otimes$ & homomorphic multiplication \\
    $\text{Rot}(\mathbf{c}, i)$ & rotation of $\mathbf{c}$ by $i$ steps & $\mathbb{I}$ & indicator function \\
    $L$ & maximum level & $l$ & current level \\
    $d$ & hidden dimension & $t$ & number of interleaved replication \\
    $n_0$ & input sequence length & $n'$ & current sequence length \\
    $N'$ & CKKS polynomial degree & $N$ & number of slots \\
    $q_i$ & $i$-th modulus of ciphertext & $\Pi$ & product of the modulus chain \\
    \bottomrule
    \end{tabular}%
\end{threeparttable}
\end{table}

\subsection{RNS-CKKS Scheme}
\label{subsec:back:ckks}
\supp{The FHE scheme used in this paper is the full residue number system (RNS) variant of Cheon-Kim-Kim-Song (CKKS)\cite{cheon2017homomorphic,cheon2018full}. $L$ represents the multiplicative depth, i.e., the maximum number of sequential multiplications supported. In RNS of $L$ levels, the large ciphertext modulus $\Pi$ is decomposed into $L + 1$ small components $\{q_0,\cdots,q_L\}$. A fresh ciphertext starts from level $L$ and drops to $L-1$ after a homomorphic multiplication and rescaling. 
When remaining level is 0, no subsequent multiplication is allowed. Then an expensive bootstrapping operation \cite{cheon2018bootstrapping} is needed to refresh the ciphertext and raises the modulus from $q_0$ to $\prod_{i=0}^L q_i$, enabling continued evaluations. We treat this operation as a blackbox. }

The CKKS scheme also supports Single Instruction Multiple Data (SIMD) encoding. Specifically, a message $m\in\mathbb{C}^N$ is transformed into a plaintext polynomial $\mathbb{Z}[X]/(X^{N'}+1)$ with a special discrete
Fourier transform (DFT), where $N'=2N$, which is then encrypted as a ciphertext. SIMD encoding enables parallel homomorphic operations across all slots. Due to the disparity of computational pattern between this SIMD operation and the tensor-based model inference, the optimization of the mapping from tensors to polynomials, frequently called packing, is required. This forms a separate design space that does not affect the accuracy of encrypted inference.


The CKKS operations used in this paper are as follows:

\begin{itemize}
    \item \textbf{Addition.} 
    Compute a ciphertext that encrypts the sum of two ciphertexts: $(\mathbf{c_1}\oplus \mathbf{c_2})[i]=\mathbf{c_1}[i]+\mathbf{c_2}[i]$.

    \item \textbf{Multiplication.} A ciphertext can be multiplied with either a ciphertext (\textbf{ct-ct} multiplication) or plaintext (\textbf{ct-pt} multiplication) at the cost of one level:
    $(\mathbf{c_1}\otimes\mathbf{c_2})[i]=\mathbf{c_1}[i]\cdot \mathbf{c_2}[i]\quad\text{or}\quad(\mathbf{c_1}\otimes\mathbf{p})[i]=\mathbf{c_1}[i]\cdot \mathbf{p}[i]$.

    \item \textbf{Rotation.} Shift each element inside a ciphertext cyclically by any steps $r$:
    $\text{Rot}(\mathbf{c}, r)[i]=\mathbf{c}[i+r]$. Note that by this definition, rotation represents left shift by default.
    
    \item \textbf{Bootstrapping.} Raise the level of a ciphertext to $L$.
    
    \item \textbf{Level drop.} Drop the level of a ciphertext.
\end{itemize}

Encryption, decryption, encoding and decoding operations are omitted because they shall be executed only once at the beginning and the end of the protocol by the client, and none of our methodologies involves them. 
 
\subsection{Generative LLM Inference}
\label{subsec:back:llm}

\begin{figure}[!tb]
    \centering
    \includegraphics[width=1.0\linewidth]{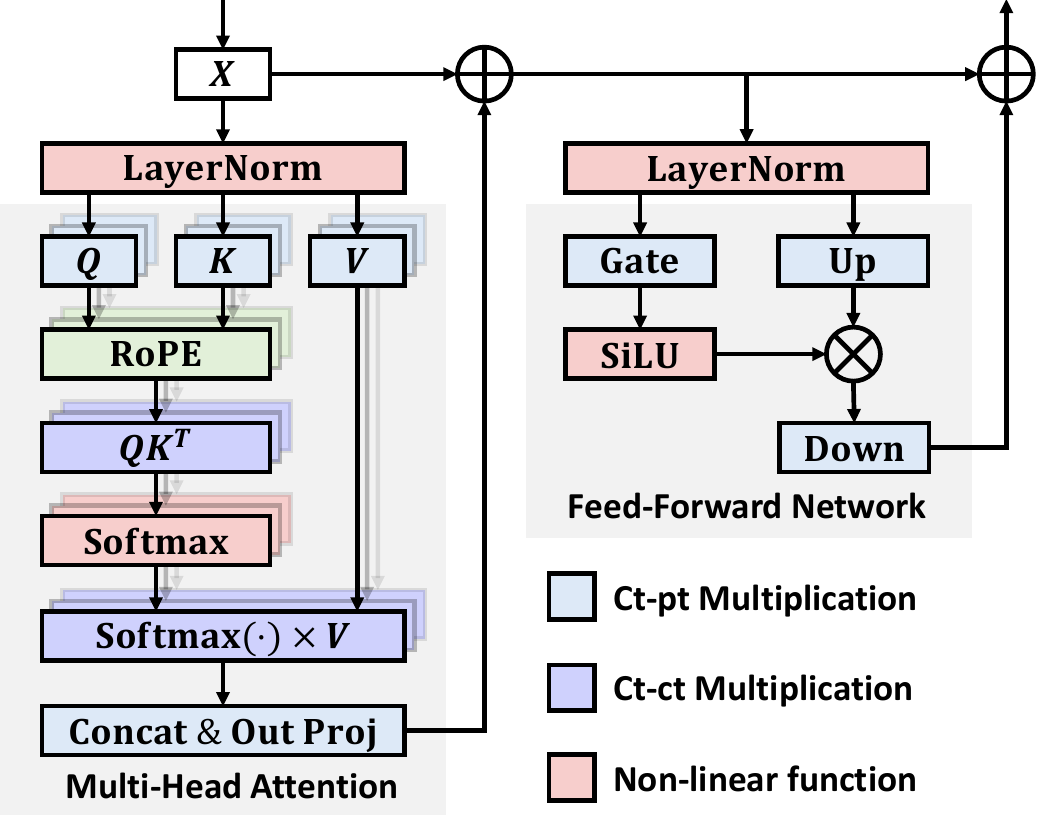}
    \caption{A Transformer block in LLaMA.}
    \label{fig:Transformer}
\end{figure}


Our work focuses on autoregressive generative Large Language Models (LLMs), which are predominantly based on a decoder-only Transformer architecture \cite{radford2019languagegpt2}. The Transformer block consists of two primary modules: a self-attention module and a Feed-Forward Network (FFN). We brief on each of them. 

\textbf{Self-Attention.} 
The input of the self-attention modules $X$ is first projected into query ($Q$), key ($K$), and value ($V$) matrices. Then, the self-attention mechanism is computed as
$Attn = \mathrm{Softmax}(\frac{QK^\top}{\sqrt{d}}) V.$
A final output projection takes $Attn$ as input and generates the output of the self-attention module. 

Generative LLM inference primarily comprises two stages, i.e., the prefill stage and the decoding stage, which involve slightly different self-attention computation. In the prefill stage, LLMs process the entire prompt sequence to compute and cache the key and value matrices for all tokens, denoted as $K_\mathrm{cache}$ and $V_\mathrm{cache}$. The decoding stage then generates output tokens sequentially based on the KV cache. At step $t$, LLM takes the most recent token $X[t-1]$ as input and compute its query, key, and value vectors, denoted as $Q_\mathrm{dec}^t, K_\mathrm{dec}^t, V_\mathrm{dec}^t$, respectively. The new key and value vectors are then appended to the existing KV cache:
\begin{equation*}
    K_\mathrm{cache}^t \leftarrow \left[\ K_{\mathrm{cache}}^{t-1} || {K_\mathrm{dec}^t}\right],
    V_\mathrm{cache}^t \leftarrow \left[\ V_{\mathrm{cache}}^{t-1} || {V_\mathrm{dec}^t}\right],
\end{equation*}
where $\left[\cdot||\cdot\right]$ denotes concatenation. The self-attention is then computed according to the following formula: 
\begin{equation*}
    Attn_\mathrm{dec} = \mathrm{Softmax}(\frac{Q_{\mathrm{dec}}K_{\mathrm{cache}}^\top}{\sqrt{d}}) V_{\mathrm{cache}}.
\end{equation*}

In SOTA LLMs, Rotary Positional Embedding (RoPE) \cite{su2023roformer} is commonly applied to the query and key vectors to incorporate relative position information. RoPE involves rotating $Q_\mathrm{dec}$ and $K_\mathrm{dec}$ by a certain angle depending on their positions before computing $Q_{\mathrm{dec}}K_{\mathrm{cache}}$, which only introduces negligible compute \cite{su2023roformer}.

\textbf{FFN.}
For LLaMA \cite{touvron2023llama} models, FFN uses the SwiGLU activation function, and is defined as
\begin{equation*}
   \mathrm{FFN}(x) = \left(\text{SiLU}(XW_\text{gate})\odot (XW_\text{up})\right)W_\text{down}, 
\end{equation*}
where $W_\mathrm{up}, W_\mathrm{gate}, W_\mathrm{down}$ are FFN parameters and $\odot$ denotes element-wise multiplication. We show the architecture of a decoder block of the LLaMA model in Figure \ref{fig:Transformer}.

\subsection{Threat Model}
We adopt the standard threat model established in prior secure inference frameworks \cite{juvekar2018gazelle,hao2022iron,Zhang2024SecureTI,rathee2020cryptflow2,lu2023bumblebee}. The model involves two parties: a server, holding a proprietary LLM in plaintext, and a client, possessing a private input. Both parties are assumed to be semi-honest, meaning they correctly execute the protocol but may attempt to learn additional information from the messages they receive. Our goal is to ensure that the server's proprietary model weights and the client's private input remain confidential from the other party. The security of our protocol reduces to the security of the underlying CKKS FHE scheme.





\section{System Overview}
\label{sec:overview}

\begin{figure}[!tb]
    \centering
    \includegraphics[width=\linewidth]{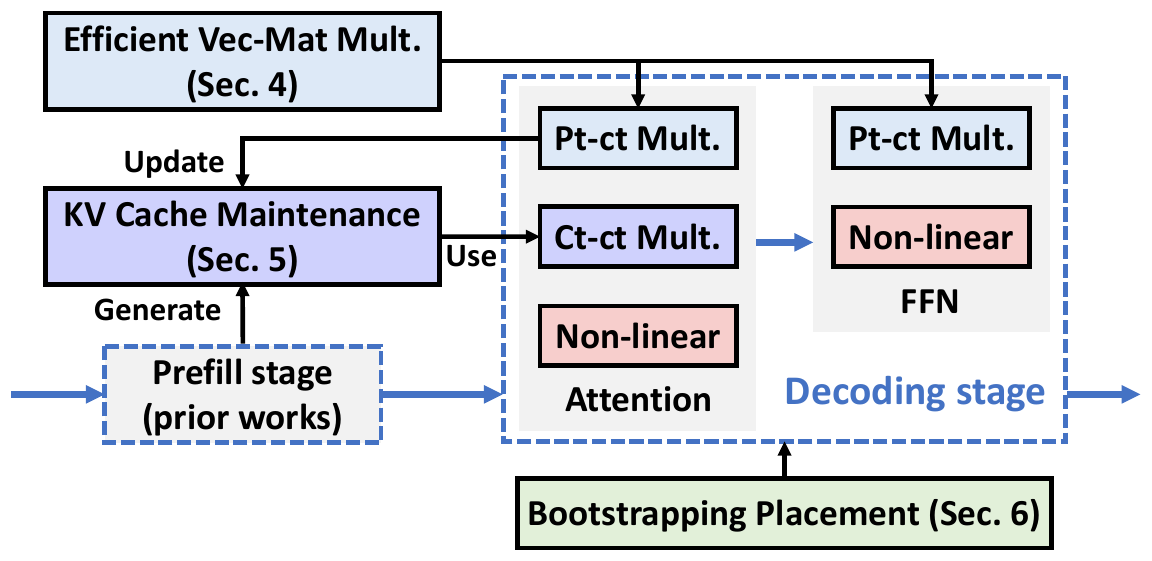}
    \caption{Overview of \method. }
    \label{fig:overview}
\end{figure}

Figure \ref{fig:overview} provides an overview of \method. To implement an efficient Transformer inference system with KV cache, our method features three components. First, we introduce an efficient ct-pt VMM algorithm with interleaved replicated packing in Section \ref{sec:gemv} to address the slot underutilization problem. Secondly, we design a series of protocols for KV cache update and computation, drastically reducing the complexity of the attention block, as explained in Section \ref{sec:kv_cache}. Third, Section \ref{sec:bootstrap} adapts the bootstrapping placement algorithm of Orion \cite{Ebel2023OrionAF} to Transformer models, fully exploiting the search space enlarged due to the activation size reduction.

\section{Efficient FHE Protocol for VMMs}
\label{sec:gemv}



\begin{figure*}[!tb]
    \centering
    \includegraphics[width=\linewidth]{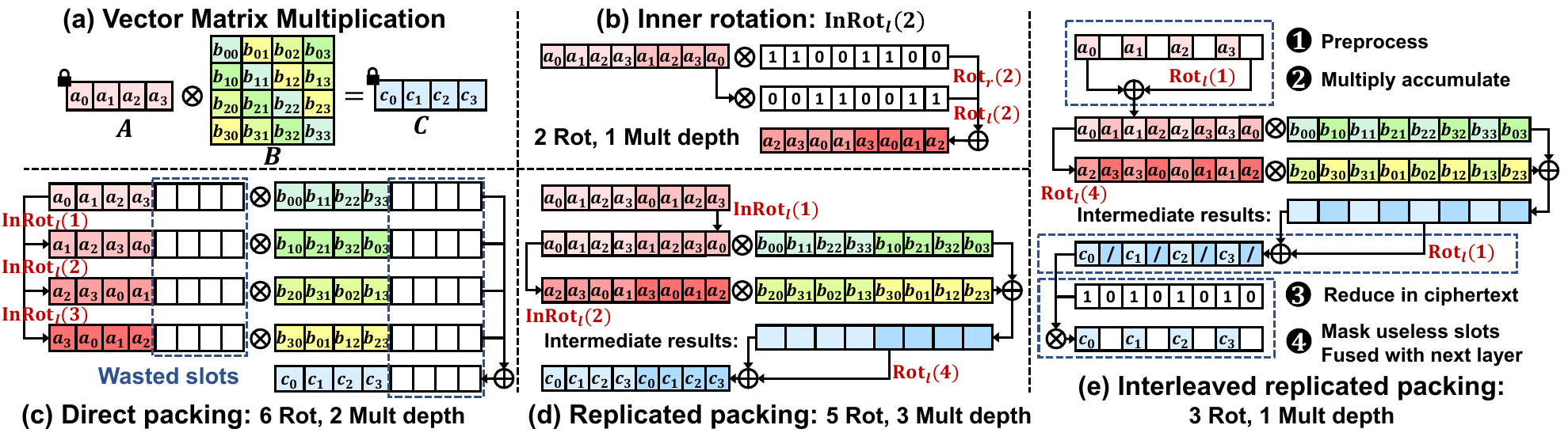}
    \caption{Toy example of our VMM packing method, where $\text{Rot}_r$ and $\text{Rot}_l$ denote right-rotate and left-rotate, respectively. (a) Ciphertext-plaintext vector matrix multiplication. (b) Example of inner rotation $\text{InRot}_l(2)$. (c) Direct packing requires 6 rotations and 2 multiplication depths. (d) Replicated packing requires 5 rotations and 3 multiplication depths. (e) Interleaved replicated packing requires 3 rotations and 1 multiplication depth, where masking useless slots can be fused with the next layer.}
    \label{fig:gemv}
\end{figure*}

\supp{In this section, we present our efficient ct-pt VMM protocol. Transformer models involve extensive VMM operations between ciphertext activations and plaintext weight matrices, including QKV generation, output projection in Attention layers, as well as up and down projections in FFNs.} 

\supp{Previous work on optimizing the ct-pt multiplication primarily focused on the \textbf{matrix-matrix multiplications in the prefill stage}, which is inefficient for the \textbf{VMMs in the KV-cache-enabled decoding stage}. For instance, in BOLT \cite{pang2023bolt}, the activation tensor for the prefill stage, with a size of $\mathbb{R}^{n\times d}$, is large enough to pack into multiple ciphertexts. However, for the decoding stage, the activations are a vector of size $\mathbb{R}^{1\times d}$, which requires zero padding to fill all $N$ slots of a ciphertext to perform the calculation correctly. NEXUS \cite{Zhang2024SecureTI} proposed an amortization-friendly offline-online computing strategy, where offline preprocessing enables the amortization of matrix multiplication costs across multiple decoding steps. However, the overhead of preprocessing is dependent only on the size of the weight matrix, not on the input dimension. Consequently, NEXUS fails to leverage the significantly lower computational complexity of the decoding stage compared to the prefill stage. Furthermore, NEXUS requires the client to perform expensive rotation (key-switch) operations, making it less deployable.
}

\supp{To address these challenges, we propose a novel packing technique called \textbf{Interleaved Replicated Packing} to enable efficient homomorphic computation. This technique effectively utilizes idle slots to reduce the number of multiplications and costly rotations. Furthermore, in Section \ref{subsec:extraction}, we fuse the masking operation for extracting valid slots with downstream layers, thereby further decreasing the consumed multiplicative depth.}

\subsection{Interleaved Replicated Packing} 
\supp{For simplicity, let's consider the plaintext weight matrix to be a square matrix first. Let the input ciphertext vector be $A\in \mathbb{R}^d$ and the plaintext weight matrix be $B\in \mathbb{R}^{d\times d}$. We need to compute $A\times B=C\in \mathbb{R}^d$. Figure \ref{fig:gemv}(a) shows a toy example for $N=8, d=4$.}

\supp{Our algorithm is based on the diagonal packing \cite{Halevi2014helib}, which computes VMMs by first extracting the generalized diagonals of $B$. Each diagonal is then multiplied by the ciphertext of $A$, which has been rotated correspondingly. The result $C$ is obtained by summing up all these partial products.}

\textbf{Direct Packing}.
We can directly utilize the diagonal method, as shown in Figure \ref{fig:gemv}(c). To correctly compute other permutations of the input vector, we need \textbf{inner rotation} (Figure \ref{fig:gemv}(b)), which costs 2 rotations, 2 ct-pt multiplications, and an extra multiplication depth. Overall, the entire computation using direct packing requires \textbf{$2d$ rotations} and \textbf{2 multiplication depths}, which is suboptimal due to wasted slots that are padded with zero, incurring the same cost while do not contributing to meaningful computation.



\textbf{Replicated Packing}.
To improve slot utilization of direct packing, we can replicate the input vector $A$ to fill all available slots, enabling simultaneous multiplication with multiple diagonals of matrix $B$. We term this approach \textbf{replicated packing}, shown in Figure \ref{fig:gemv}(d). However, this method requires extra inner rotations to align the replicated input vectors with their corresponding plaintext matrix diagonals. This method requires \textbf{$O(2d^2/N+2N/d)$ rotations} and consumes \textbf{3 multiplication depths}. Although it is more efficient than direct packing in slot utilization, the reliance on inner rotations results in a higher multiplicative depth overhead.



\textbf{Interleaved Replicated Packing.} Building on the insights from the limitations of direct and replicated packing, 
propose our \textbf{interleaved replicated packing} method, as shown in Figure \ref{fig:gemv}(e). In contrast to the interleaved batching method of MOAI\cite{zhang2025moai} that aims to simultaneously process multiple queries, our method aims to achieve the high slot utilization of replicated packing while eliminating the expensive inner rotation operations.

The key innovation lies in packing the input vector $A$ and the weight matrix $B$ in an interleaved manner. Interleaved packing refers to a data layout where vector elements are spaced across the ciphertext slots, such as $[a_0,0,a_1,0,a_2,0,a_3,0]$. This allows us to achieve the same effect as replicated packing using only standard rotation operations, removing expensive inner rotations.

\supp{Figure 4(e) illustrates our method, which involves four steps:
\ding{182} \textbf{Preprocess.} Similar to replicated packing, we first need to permute the replicated vectors to align with different weight diagonals. This is achieved with $\log(N/d)$ rotations and add operations, resulting in the interleaved replicated input ciphertext, e.g., $[a_0,a_1,a_1,a_2,a_2,a_3,a_3,a_0]$.
\ding{183} \textbf{Multiply accumulate.} Unlike direct replicated packing, our interleaved packing allows the permutation of all packed vectors in one ciphertext with $O(d^2/N)$ rotations instead of inner rotations, in order to align with the subsequent weight diagonals.
\ding{184}~\textbf{Reduce in ciphertext.} After step 2, we obtain intermediate results corresponding to different groups of plaintext diagonals. These results are also arranged in an interleaved fashion within a single ciphertext. We then use $\log(N/d)$ rotations and add operations to align and accumulate these intermediate results.
\ding{185} \textbf{Mask useless slots.} The final result obtained in step~3 contains only $d$ slots with valid information. 
To ensure correctness for subsequent computations, we apply an additional masking operation to zero out invalid slots and extract valid slots. Although this step introduces an extra multiplication depth, in Section \ref{subsec:extraction} we show how this masking operation can be fused with other operations in the Transformer model, thereby avoiding consuming an extra level.}

\supp{Our method completely fills the available slots, avoids the need for inner rotations, and significantly reduces the computational overhead. Overall, interleaved replicated packing method requires only \textbf{$O(2\log(N/d)+d^2/N)$ rotations and a multiplication depth of 1}. The complexity can be further optimized by the BSGS method. } 

\supp{\textbf{BSGS Optimization.} The Baby-Step Giant-Step (BSGS) \cite{ju2024neujeans,pang2023bolt,xu2024privcirnet} algorithm is a widely used method for reducing the number of rotations. It divides the rotation process into a baby-step phase before multiplication and a giant-step phase after multiplication.} \supp{For the direct packing and replicated packing, we can fuse the inner rotation of giant-step with the multiplication by the matrix $B$, reducing the rotation complexities to $O(4\sqrt{d})$ for direct packing and $O(4\sqrt{\frac{d^2}{N}} + 4\sqrt{\frac{N}{d}})$ for replicated packing, without increasing the multiplicative depth.} \supp{For interleaved replicated packing, BSGS can be applied directly to step 2, which reduces the rotation complexity to $O(2\log(N/d)+2\sqrt{d^2/N})$.}


\supp{\textbf{General Case for Non-square Matrices.} }
We then generalize the method to the case of padded non-square weight matrices, namely $B\in\mathbb{R}^{d\times \alpha d}$ or $B\in\mathbb{R}^{\alpha d\times d}$, where $k$ is a power of two. Due to space limitations, we only briefly discuss the key modification here, with further implementation details of interleaved replicated packing provided in Appendix \ref{apd:gemv}.

Intuitively, in the interleaved replicated packing algorithm, the lengths of the input and output vectors are controlled by Step 1 and 4, respectively. If the two steps share the same number of operations, then the length remains the same, as shown in the example above. However, this symmetry is not necessary: if we reduce more slots than the preprocessing step has replicated, the vector is shortened or vice versa. For weight matrices, we can keep the diagonal packing pattern, while extending the diagonals into those of their square submatrices. With such modifications, we can complete the general case with the number of rotations being $O(\log(\alpha N^2/d^2)+2\sqrt{\alpha d^2/N})$.

\supp{\textbf{Theoretical complexity analysis.}
Table \ref{tab:gemv_new} shows the complexity comparison of our interleaved replicated packing with prior-art methods in the number of rotations and pt-ct multiplications. Our technique saves about $2521\times$ and $7.2\times$ rotations compared to NEXUS and BOLT.}

\begin{table}[!tb]
\caption{Theoretical complexity comparison of ciphertext-plaintext multiplication. We show the case where weight matrix is a square matrix. The concrete numbers are based on the dimensions of Q,K,V generation of Llama-3-8B, with $(N, d)=(32768, 4096)$. BOLT+padding means padding the ciphertext vector to a size of $N$ to correctly utilize the BSGS method. NEXUS's amortization factor, denoted as $n$ (the number of decoding steps), is set to 256.}
\footnotesize
\label{tab:gemv_new}
\resizebox{0.9\linewidth}{!}{
\begin{tabular}{@{}c@{}|@{}cc@{}c@{}}
\toprule
Method & \#pt-ct Mult. & \#Rotation & Mult. Depth \\
\midrule
BOLT \cite{pang2023bolt} & $O(d)$ & $O(2\sqrt{N})$ & \multirow{2}{*}{1} \\
+padding & 4096 & 375 & \\
\midrule
\multirow{2}{*}{NEXUS\cite{Zhang2024SecureTI}}  & $O(\frac{d^2}{n})$ & $O(\frac{2d^2}{n})$ & \multirow{2}{*}{1} \\
 & 65536 & 131072 & \\
\midrule
Direct  & $O(2d+\sqrt{d})$ & $O(4\sqrt{d})$ & \multirow{2}{*}{2} \\
w/ BSGS & 8254 & 252 & \\
\midrule
Replicated  & $O(\frac{2d^2}{N}+\sqrt{\frac{d^2}{N}} + 4\sqrt{\frac{N}{d}})$ & $O(4\sqrt{\frac{d^2}{N}} + 4\sqrt{\frac{N}{d}} )$ & \multirow{2}{*}{3} \\
w/ BSGS & 1046 & 103 & \\
\midrule
Ours  & $O(\frac{d^2}{N})$ & $O(2\sqrt{\frac{d^2}{N}} + 2\log \frac{N}{d})$ & \multirow{2}{*}{1} \\
w/ BSGS & 512 & 52 & \\
\bottomrule
\end{tabular}
}
\end{table}


\subsection{Fused Ciphertext Extraction}
\label{subsec:extraction}

We apply the interleaved replicated packing algorithm to all ct-pt VMMs in each decoder block.
During inference, useless slots need to be cleaned up after each of the multiplications to avoid disrupting subsequent layers. Directly multiplying the output with another boolean mask would incur additional level consumption and raise the levels of many operations, as well as severely increase the bootstrapping latency. To address this, we propose a customized method that eliminates the additional level of consumption. 

The key observation is that despite the existence of useless slots, the number of ciphertexts never increases after any ct-pt multiplications, as $t\cdot d\equiv N$. Therefore, we can defer the cleaning process without additional overhead, and merge it with the next layer, in which element-wise multiplication is always involved. 

We elaborate this idea by the extraction after Q, K generation. In typical models like Llama \cite{touvron2023llama}, the $Q$ and $K$ matrices need to go through RoPE \cite{su2023roformer}, which is naturally realized by element-wise multiplication. Thus, we can extract the useful slots easily by setting the coefficients of other slots to zero:
\[
\mathbf{p}^{\text{RoPE}}[i]=
\cos (\theta(i, n))\mathbb{I}\{t\mid i\}
\]
where $n$ is the index of token. We omit other two plaintexts encoding sine information, which is essentially the same. After the SIMD multiplication, we compute RoPE by rotation and addition. Detailed algorithm is given in Appendix \ref{apd:RoPE}.

For other linear layers using our interleaved replicated packing, exactly the same method can be applied. After up and gate projection, out and down projection, and V generation, the mask may be merged with SiLU function, layer normalization, and the mask for updating V cache, respectively.

\textbf{Level consumption analysis.}
With the tailored methods above, we have efficiently exempted all level consumption incurred by ciphertext extractions, reducing the overall level consumption of ct-pt multiplication from 10 to 5. 

\section{KV Cache Maintenance} 
\label{sec:kv_cache}

\supp{
In this section, we introduce our encoding method for handling the KV cache. During the decoding process, all ct-pt multiplications adopt the interleaved replicated packing method described in Section \ref{sec:gemv}. Therefore, the input encoding for the Attention block must also be compatible with the scheme. This section first explains the computation of ct-ct VMMs for $QK^T$ and $\text{Softmax} \times V$, as well as the arrangement and update of the KV cache matrix during the decoding stage. Then we detail the generation of the KV cache. For simplicity, we first consider the case where the number of heads $H=1$ in the MHA.}

\begin{figure}[!tb]
    \centering
    \includegraphics[width=\linewidth]{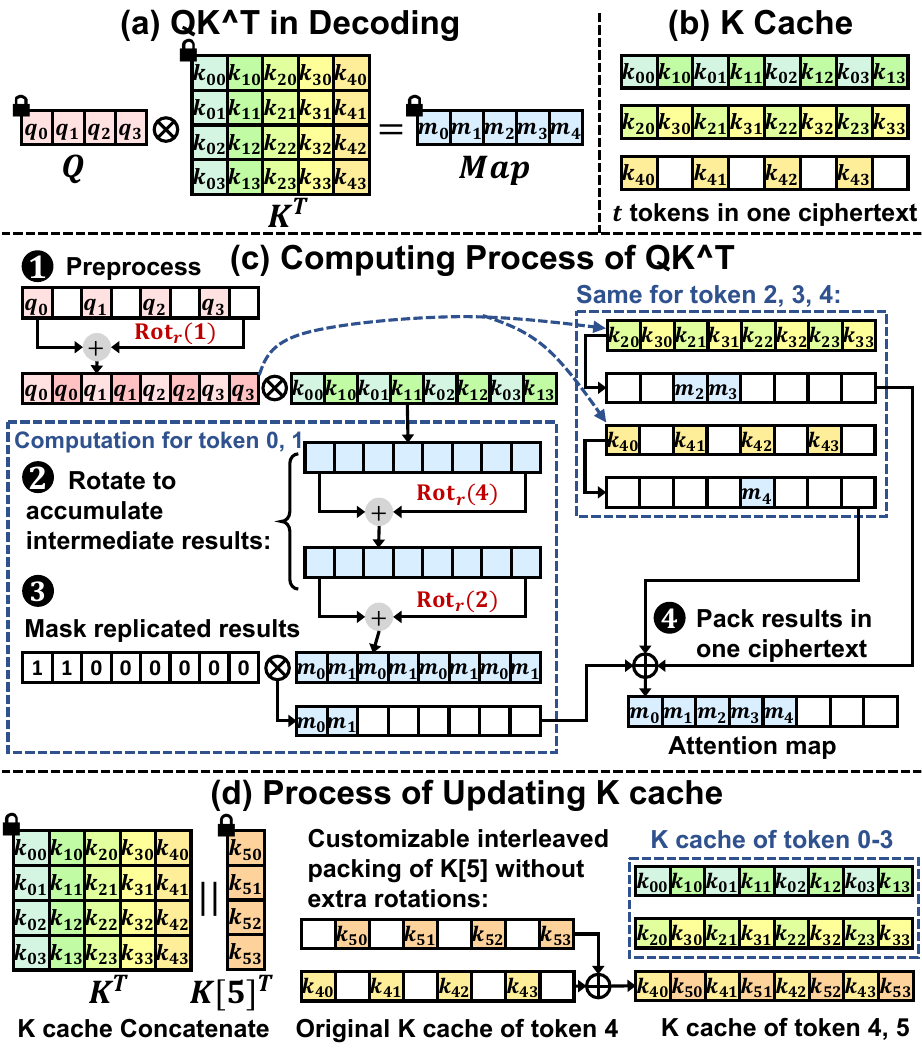}
    \caption{Example of $QK^T$ and K Cache Maintenance. (a) $QK^T$ in decoding, where existing sequence length $n'=5$. (b) Interleaved packing for K Cache, where K Cache of $t=N/d=2$ tokens are packed in one ciphertext. (c) Computing process of $QK^T$. (d) Process of updating K cache when generating token 5.}
    \label{fig:k_cache}
\end{figure}

\subsection{$QK^T$ and K Cache Maintenance}
\supp{First, let's consider the computation of $QK^T$ within the Attention module during the decoding stage. Assume the current number of tokens is $n'$, and the KV cache already holds information for these tokens. At this point, we have $Q\in \mathbb{R}^{1\times d}$ and $K^T\in \mathbb{R}^{d\times n'}$. We need to compute the attention map, i.e., $Map=QK^T\in \mathbb{R}^{1\times n'}$. Figure \ref{fig:k_cache}(a) shows a toy example where $N=8, d=4$ and $n'=5$.}

\supp{\textbf{Interleaved Packing of K Cache.} 
Since the computation of $Q=XW_Q$ uses the interleaved replicated packing from Section \ref{sec:gemv}, the resulting $Q$ vector is also packed in an interleaved format. To match this packing, we adopt a similar strategy for the K cache.} \supp{For a vector $Q$ with dimension $d$, we can replicate it within a single ciphertext by $t=N/d$ times. We pack the K caches for \textbf{$t$ tokens into a single ciphertext}. In the example shown in Figure \ref{fig:k_cache}(b), where $t=2$, we pack the K cache of token 0 and token 1 into one ciphertext in an interleaved manner:  $[k_{00},k_{10},k_{01},k_{11},k_{02},k_{12},k_{03},k_{13}]$, where $k_{ij}$ denotes the $j$-th K cache dimension of the $i$-th token. 
The total number of ciphertexts of K cache is  $\lceil n'/t \rceil$. This compact packing method allows for fully utilizing the ciphertext slots and minimizing the storage for the K cache.}

\textbf{$QK^T$ Computation.}
When computing $QK^T$, both $Q$ and $K$ are ciphertexts, necessitating expensive ct-ct multiplications with key-switch operations. Here, we again leverage interleaved replicated packing to make full use of the ciphertext slots. Figure \ref{fig:k_cache}(c) illustrates our method, which involves four steps: 
\ding{182} \textbf{Preprocess.} The query vector $Q$ is packed in interleaved format with unused slots. To create the desired replicated input for parallel computation, we perform $\log (N/d)$ rotations to replicate the vector (e.g., $[q_0,q_0,q_1,q_1,q_2,q_2,q_3,q_3]$), which is similar to the pre-processing step in Section \ref{sec:gemv},
\ding{183} \textbf{Multiply and accumulate intermediate results.} The preprocessed $Q$ ciphertext is then multiplied by the interleaved K cache ciphertext. We accumulate these partial results using $\log d$ rotations and additions, obtaining a replicated attention map $[m_0,m_1,m_0,m_1,m_0,m_1,m_0,m_1]$, where $m_i=\sum\limits_jq_jk_{ij}$.
\ding{184} \textbf{Mask replicated results.} We multiply the results by a binary mask (e.g., $[1,1,0,0,0,0,0,0]$) to extract a single valid copy of the replicated attention scores. The same process (including Step 2 and 3) is conducted for the other ciphertexts of K Cache to compute $m_2, m_3$, and $m_4$.
\ding{185} \textbf{Pack results in one ciphertext.} Finally, the masked results from the different token groups are accumulated to pack the full attention map into a single ciphertext $[m_0,m_1,m_2,m_3,m_4,0,0,0]$. This compact packing significantly reduces the overhead for subsequent operations like Softmax and bootstrapping. Overall, the computation of $QK^T$ requires $O(\lceil n'/t \rceil)$ ct-ct multiplications and $O(\lceil n'/N \rceil\log(N/d)+\lceil n'/t \rceil \log d)$ rotations.

\textbf{K Cache Update.} Our packing scheme for K cache not only enables efficient ciphertext slot usage for fast multiplication, but also supports rapid K Cache updates. As illustrated in Figure \ref{fig:k_cache}(d), before generating token 6, we need to add the $K$ vector of the previously generated token 5 to the K Cache. Similar to $Q$, the result of $K[5]=X[5]W_K$ is also in an interleaved packing format, which is consistent with our K Cache packing scheme. 

If the last ciphertext of the K Cache still has available slots (i.e., current length $n'$ is not a multiple of $t$, as in Figure \ref{fig:k_cache}(b)), we align $K[5]$ with the empty slots by generating \textbf{customizable interleaved packing}. For instance, by changing $\text{Rot}_l(1)$ to $\text{Rot}_r(1)$ in Step 3 in Figure \ref{fig:gemv}, the valid accumulated result is shifted to the next position, producing a ciphertext like $[0,k_{50},0,k_{51},0,k_{52},0,k_{53}]$. If $n'$ is a multiple of $t$, the new $K$ can simply serve as a new ciphertext for the K Cache. Ultimately, this approach allows the K Cache to be updated at \textbf{negligible cost}.

\subsection{$\text{Softmax}\times V$ and V Cache Maintenance}

\begin{figure}[!tb]
    \centering
    \includegraphics[width=\linewidth]{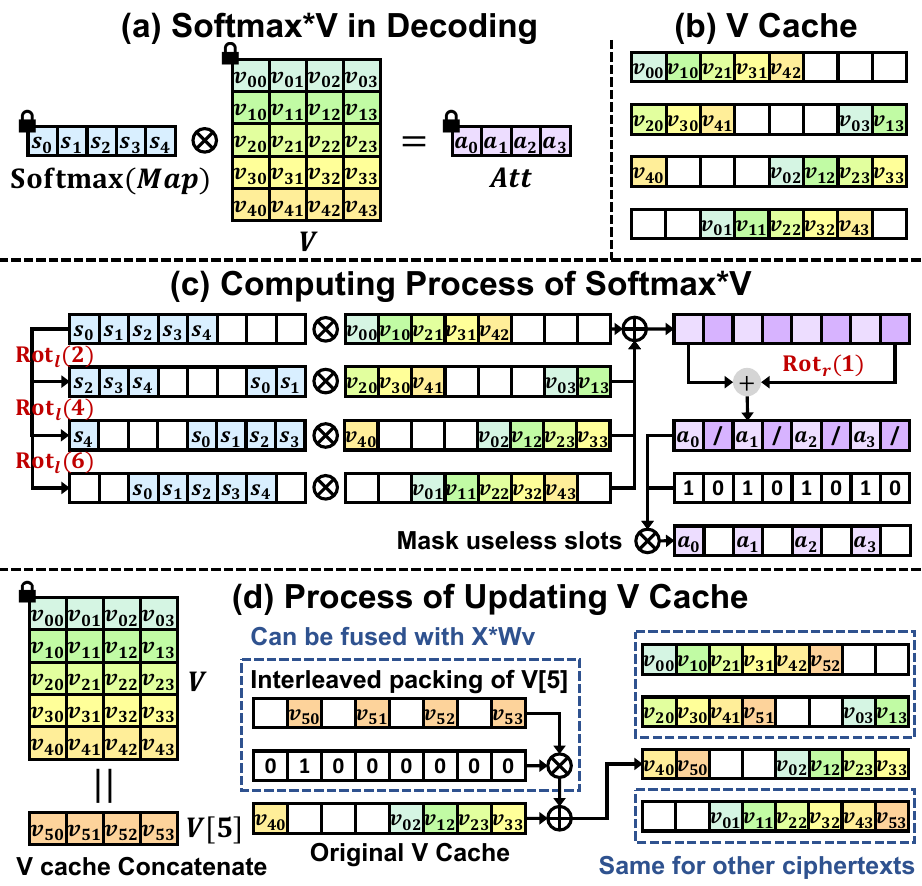}
    \caption{Example of our $\text{Softmax}\times V$ and V Cache Maintenance. (a) $\text{Softmax}\times V$ in decoding, where existing sequence length $n'=5$. (b) Example of packing for V Cache. (c) Computing process of $\text{Softmax}\times V$. (d) Process of updating V cache when generating token 5.}
    \label{fig:v_cache}
\end{figure}

\supp{Now consider the second ct-ct multiplication operation in the Attention module. For $\text{Softmax}(Map)\in\mathbb{R}^{1\times n'}$ and the existing V Cache $V\in \mathbb{R}^{n'\times d}$, we need to compute $Att=\text{Softmax}(Map)V\in \mathbb{R}^{1\times d}$. Figure \ref{fig:v_cache}(a) shows an example where $N=8, d=4$ and $n'=5$.}

\supp{\textbf{V Cache packing and Computing Process}. To minimize the number of ciphertexts for Softmax and bootstrapping, the Attention Map is tightly packed into one ciphertext during the  $QK^T$ computation. In this case, one ciphertext can store attention map for at most $N$ tokens (for the case of $H$=1) with a contiguous packing scheme. We need to design a V Cache packing that enables fast computation of $\text{Softmax}\times V$ and output the result in \textbf{interleaved} format, while also facilitating easy updates for V Cache. For every $N$ tokens, we require $d$ ciphertexts to store their V Cache.}

\supp{Naive diagonal packing (e.g., $\{[v_{00},v_{11},...],[v_{10},v_{21},...],...\}$), would get result in contiguous format, which is unsuitable for the subsequent calculation. Instead, we make the $i$-th element of each ciphertext belong to the $\lfloor i/t \rfloor$ column of the V Cache, as shown in Figure \ref{fig:v_cache}(b). This ensures that every $t$ elements across all ciphertexts belong to the same column, ensuring that the result $Att$ form an interleaved packing with an interval of $t$. This allows us to directly use the method described in Section \ref{sec:gemv} for the next  multiplication.}

\supp{Our packing method enables efficient computation of $\text{Softmax}\times V$, as shown in Figure \ref{fig:v_cache}(c). We compute intermediate results via $d-1$ rotations and $d$ multiplications on the Softmax result ciphertext. Then, apply $\log(N/d)$ rotations and additions within the intermediate result ciphertexts. A masked multiplication then yields $Att$ in interleaved form. This process requires a multiplicative depth of 2. By further utilizing the BSGS method and Multi-Head Optimization, the total rotations are reduced to $O((2\sqrt{2d}+\log(N/d))\cdot \lceil n'/N\rceil))$.}

\supp{\textbf{V Cache Update.} In our packing, each token’s $V$ vector is distributed across multiple ciphertexts \textbf{in an interleaved manner}. This design allows for efficient V Cache updating. As shown in Figure \ref{fig:v_cache}(d), to add $V[5]$, we can generate a customizable interleaved packing of $V[5]$ ($[0,v_{50},0,v_{51},0,v_{52},0,v_{53}]$), which is similar to the method for the K Cache. Through $d$ masked multiplications, we can then quickly add $V[5]$ to the V Cache. Moreover, as described in Section \ref{subsec:extraction}, we fuse the masking operation during the update process when generating $V[5]$ (Figure \ref{fig:gemv}, step 4), avoiding any extra multiplicative depth. Thus, the V Cache can be updated with minimal overhead using inexpensive pt-ct multiplications.}

\subsection{Efficient Multi-head Attention}
\supp{We now extend our method to multi-head attention (MHA) with $H>1$. Following prior work \cite{xu2025blb}, we exploit the independence between heads to reduce overhead. Our packing strategy interleaves elements from different heads within a single ciphertext. The smallest unit for this interleaving is a block of $t=N/d$ elements. This allows rotation operations to process multiple heads in parallel.}

\supp{For $QK^T$ computation shown in Figure \ref{fig:k_cache}, we consider $d=4$ and $H=2$. We partition $K$ into heads: $K^{h=0} = K[:,0:1]$, $K^{h=1} = K[:,2:3]$. We pack $t$ tokens into a single ciphertext, and the hidden dimensions are reordered by head: e.g., $[k_{00},k_{10}|k_{02},k_{12}|k_{01},k_{11}|k_{03},k_{13}]$. The packing order for $Q$ is modified accordingly. This packing arrangement for $Q$ and $K$ is achieved by reordering the plaintext $W_Q$ and $W_K$. The resulting $QK^T$ yields attention maps for all heads in one ciphertext: e.g., $[m_0^{h=0}, m_1^{h=0} | m_0^{h=1}, m_1^{h=1} | 0,0,0,0]$, where one ciphertext can pack up to $N/H$ token attention map results. }

\supp{A similar multi-head optimization is applied to the $\text{Softmax}\times V$ computation. Since the $\text{Softmax}(Map)$ can pack up to $N/H$ token results, we require $d/H$ ciphertexts to pack the V cache for every $N/H$ tokens. The final $Att$ results are also head-interleaved: $[a_0^{h=0},0|a_0^{h=1},0|a_1^{h=0},0|a_1^{h=1},0]$.}

\supp{These optimizations reduce the number of rotations to $O(\lceil n'/N \rceil \cdot\log(N/d)+\lceil n'/t \rceil \cdot\log(d/H))$ for $QK^T$ and $O((2\sqrt{2d/H}+\log(N/d))\cdot \lceil n'/N\rceil))$ for $\text{Softmax}\times V$. Although $n'$ appears in the expressions, in most cases, a single ciphertext gives enough context length, being able to contain 1024 tokens for Llama-3-8B, and more for smaller models like TinyLlama or GPT2-base.}

\supp{For the final $\text{Concat}(Att_h)W_O$ , we do not require any repacking for concatenation. Instead, we can directly compute the output projection by reordering the plaintext weight $W_O$.}

\subsection{KV Cache Generation}
The inference flow above is also applicable to the prefilling stage with several minor modifications. In ciphertext-plaintext multiplications, a ciphertext can naturally contain $t$ tokens without any preprocessing. In $QK^T$, we execute additional inner rotations to enable multiplications between vectors with different indices in the $t$ dimension. During the generation of V cache, multiplications with mask may be processed in a SIMD manner with batch size $t$. Admittedly, this prefilling method is suboptimal due to additional constraints on the form of $K$ and $V$, it turns out that the overhead amortized across generated tokens is negligible compared to the benefit of adopting KV cache in Section \ref{sec:eval}.

\section{Efficient Bootstrapping Placement}
\label{sec:bootstrap}
Bootstrapping placement has a critical impact on the end-to-end latency of FHE-based inference, because it directly determines the total latency of all bootstrappings and the input level of each module. As the shapes and sizes of activation tensors keep changing in the inference process, prior works \cite{Zhang2024SecureTI,moon2024thor,zhang2025moai} place bootstrappings for FHE-based Transformer inference manually and only insert the bootstrapping when the activation tensor size is small. 
Nevertheless, this rule no longer holds in the presence of the KV cache, as we can always hold the full activation in one ciphertext. 


The SOTA bootstrapping placement algorithm, Orion\cite{Ebel2023OrionAF}, formulates bootstrapping placement as a shortest-path problem over a Directed Acyclic Graph (DAG) derived from the network structure. 
While effective for shallow convolutional models, Orion encounters two fundamental limitations when applied to large Transformer architectures. First, its time complexity, $\mathcal{O}(L^2 d_1 d_2)$, becomes prohibitive for deep models, where $L, d_1, d_2$ represents the maximum level, the number of layers in one decoder, and the number of decoders, respectively. Second, Orion restricts bootstrapping to layer boundaries and excludes placements inside non-linear modules. Ignoring such in-module placements often leads to suboptimal latency. In this section, we address both issues through DAG simplification and in-module bootstrapping.



\begin{figure}[!tb]
    \centering
    \includegraphics[width=\linewidth]{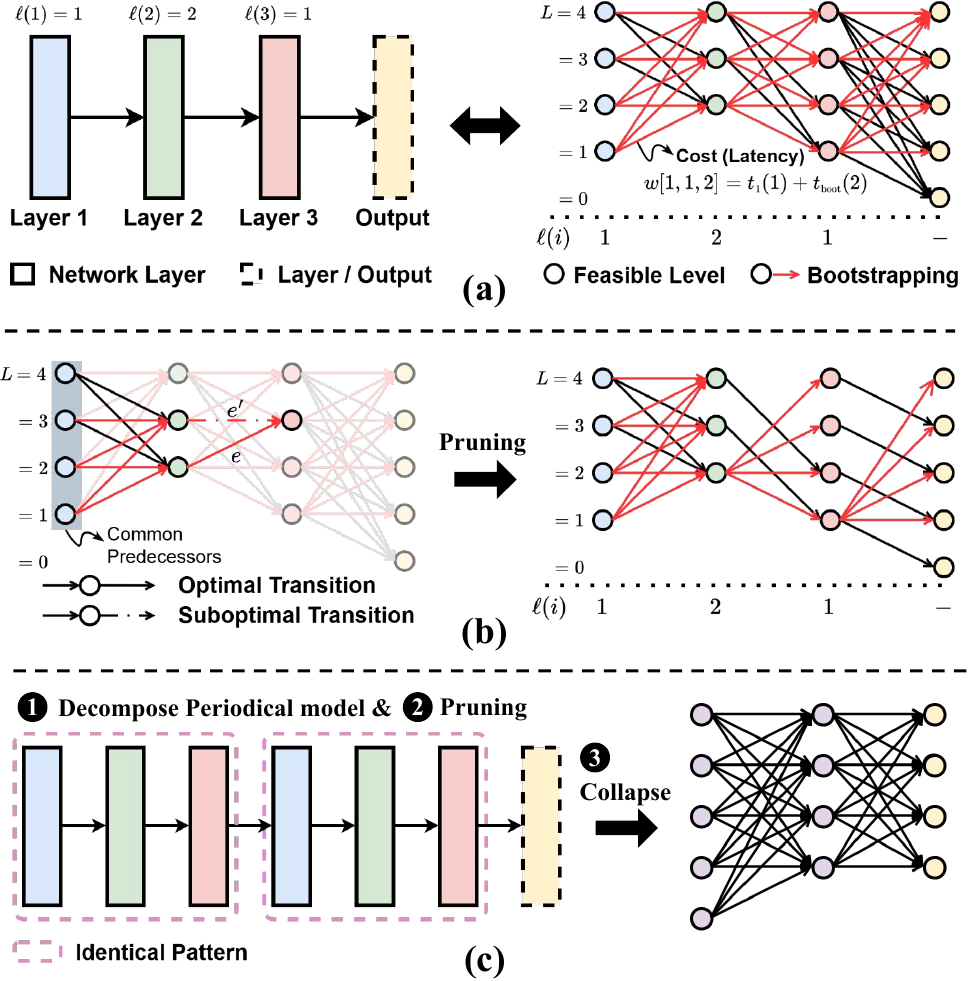}
    \caption{
    Level management in \method. (a) Construction of the level graph. Unlike Orion which assigns computation costs to nodes, we integrate layer computation latency directly into the edge weights (defined in Equation~\ref{eqn:latency-edge}). $\ell(i)$ indicates the multiplicative depth. (b) Graph pruning via path dominance. Suboptimal transitions are identified and pruned (left) to generate a sparse level graph (right). (c) Hierarchical optimization for periodical models. 
    } 
    \label{fig:DAG}
\end{figure}


As illustrated in Figure \ref{fig:DAG} (a), Orion constructs a DAG by mapping each valid input level of every layer to a vertex, and connecting the vertices of adjacent layers with edges. Each edge denotes a feasible transition between an input–output level pair for the corresponding layer. Following Orion, we assume that layer computation is performed before any optional bootstrapping or level drop. Under this rule, input levels smaller than the multiplicative depth of a layer are invalid. Consequently, identifying the shortest path in this graph is equivalent to determining a globally optimal level assignment for all layers, thereby determining bootstrapping placement.

Formally, for a network with $D$ layers and a maximum level $L$, we denote by $v[i,j]$ the vertex corresponding to the $i$-th layer with input level $j$. Let $e[i, x, y]$ be the directed edge from $v[i, x]$ to $v[i+1, y]$ with weight $ w[i, x, y]$. In addition, we denote $\ell(i)\leq L$ as the multiplicative depth for layer $i$. Then, the minimum latency should be $\min\limits_{\mathbf{x}}\sum\limits_{i=1}^{D}w[i,x_i,x_{i+1}]$. 

\textbf{Eliminating suboptimal edges.} 
Given that the latency of level drop is negligible , and that bootstrapping performance is insensitive to the input level, we can express the weight as:

\begin{equation}
\label{eqn:latency-edge}
    w[i,x,y]\approx t_{i}(x)+1_{x-\ell(i)<y}\cdot t_{boot}(y), 
\end{equation}

where $t_i(\cdot)$ and $t_{\text{boot}}(\cdot)$ are both monotonically increasing functions, and $1_{x-\ell(i)<y}$ denotes an indicator function that equals $1$ if a bootstrapping is required (i.e., $x-\ell(i)<y$), and $0$ otherwise. In Figure~\ref{fig:DAG}(a), all red edges correspond to transitions with $1_{x-\ell(i)<y}=1$. 


Under this formulation, a large fraction of edges can be proven to be dominated and therefore never appear on a shortest path. Specifically, for $i>1$, any edge $e[i,x,y]$ satisfying $x(x - y - \ell(i)) \neq 0$ is suboptimal, since for any predecessor level $u$,
\begin{align*}
w[i-1, u, x]+w[i,x,y]&>w[i-1,u,x-1]+w[i,x-1,y], 
\end{align*}
This inequality follows directly from a term-wise comparison of the corresponding weights.

Intuitively, this states two rules: that level drop is only reasonable at source vertices, and that bootstrapping should only be executed when the level reaches zero. Figure~\ref{fig:DAG}(b) illustrates the pruning of dominated transitions based on edge-wise cost comparison under shared predecessors. The rule pertains even in the presence of shortcuts, which is processed recursively as a subgraph with its own source vertex. Removing these suboptimal edges yields a sparser level graph (right part of (b)). This reduces the number of edges from $\mathcal{O}(L^2d_1d_2)$ to $\mathcal{O}(Ld_1d_2)$.

\textbf{Decomposing periodical model.}
Moreover, as transformer models consist of dozens of identical decoders, for those with a larger decoder number, we can decompose the search process into two stages: searching for the minimum latency of one decoder, and searching for the minimum latency of the whole model, as shown in Figure \ref{fig:DAG} (c). This further reduces the complexity to $\mathcal{O}(L^2(d_1+d_2))$, where the square term of $L$ arises because topological sorting and relaxation is only applicable to single-source problems. 



\textbf{In-module bootstrapping.}
We also refine the algorithm by expanding the search space and therefore enhance the optimization result. In the original algorithm of Orion, the possibility of  bootstrapping during non-linear evaluation is overlooked because CNN only have ReLU as non-linear function. However, non-linear layers in transformer models are approximated with various approaches, which provides a large additional search space and requires a more detailed analysis to optimize. 

We implement the non-linear layers following THOR, as elaborated in Appendix \ref{apd:nonlinear}. We approximate the SiLU and GeLU function with composite polynomials, exponentiation with a single polynomial, and inversion and inverse square root with Goldschmidt iteration. We decompose the modules by tracing the number of ciphertexts. During the processes of polynomial evaluation and Goldschmidt iteration, the number of ciphertexts and consequently the bootstrapping cost surges, for which we can safely exclude the possibility of bootstrapping and group them as a whole sub-layer. On the other hand, during processes where the number of ciphertexts remains one, variance computation, it is crucial to search operation by operation. In this way, we can decompose the non-linear layers into a sequence of sub-layers, enabling us to search in a more fine-grained manner and achieve a better global runtime. Also, considering the large multiplicative depth of those layers, enabling in-module bootstrapping also reduces the maximum level, contributing to more efficient cryptographic parameters.

\section{Experimental Results}
\label{sec:eval}

\subsection{Experimental Setup}
\label{ssec:eval-step}
\textbf{Implementation and Measurement Methodology.}
\method~ is implemented with Lattigo~\cite{lattigo} as the CPU backend for CKKS and a customized Phantom library~\cite{yang2024phantom} for GPU acceleration. All experiments are conducted on a server equipped with an Intel Xeon Platinum 8558P CPU (a maximum frequency of 4 GHz, 192 threads) and a NVIDIA A100-80GB GPU.  We measure the latency of our methods in the granularity of modules to get a precise estimate of end-to-end runtime. Due to the difference in library and scale, we compute the latency of baselines using the theoretical number of operations and the average measured operation latency.

\textbf{Cryptographic Configuration.} Throughout the experiments, we set the polynomial degree to $N'=2^{16}$ and the number of slots to $N=2^{15}$ with a 1763-bit ciphertext modulus, which achieves 128-bit security according to the Homomorphic Encryption Standard~\cite{albrecht2022homomorphic}. Following THOR~\cite{moon2024thor}, we set the maximum level to $L=13$ and the multiplicative depth of the bootstrapping circuit to $K=15$. Specifically, we adopt the Coefficients-to-Slots (CtS)-first bootstrapping variant, with a CtS depth of 4 and a Slots-to-Coefficients (StC) depth of 3. To support 8-depth approximate modular reduction via a sparse secret key (Hamming weight 192), we apply sparse secret encapsulation~\cite{bossuat2022bootstrapping} to preserve 128-bit security. The RNS modulus chain is configured with $\log_2{q_0}\approx 53$ and $\log_2{q_i}\approx 41$ for all $i\geq1$ to ensure enough noise budget.

\textbf{Models, datasets and baseline.} We select three typical generative language models for the experiments, namely GPT2-base \cite{radford2019languagegpt2}, TinyLlama-1.1B \cite{zhang2024tinyllama}, and Llama-3-8B \cite{grattafiori2024llama3herdmodels}. Basic information of these models is shown in Table \ref{tab:model_info}.
The input sequence length and total sequence length varies across experiments. We evaluate the data range and model performance using Llama-3-8B, which is most sensitive to the error term introduced by approximation. We did not finetune the model on downstream tasks. We choose two datasets: we include MRPC to enable a clear comparison with prior works, and also include Wikitext-2 to illustrate the feasibility of applying the same approximation to generation tasks.

We compare our work to a series of prior works on FHE-based Transformer inference, including NEXUS, THOR, and MOAI. For both NEXUS and MOAI, we use a natural batch, which is the sequence length and the number of slots divided by the sequence length, respectively. THOR does not mention batch processing in its paper; therefore, we set the batch size to 1. Our method can also benefit from batching, but does not depend on it, and hence we also set our batch size to 1.

\begin{figure}[!t]
  \centering
  \includegraphics[width=\linewidth]{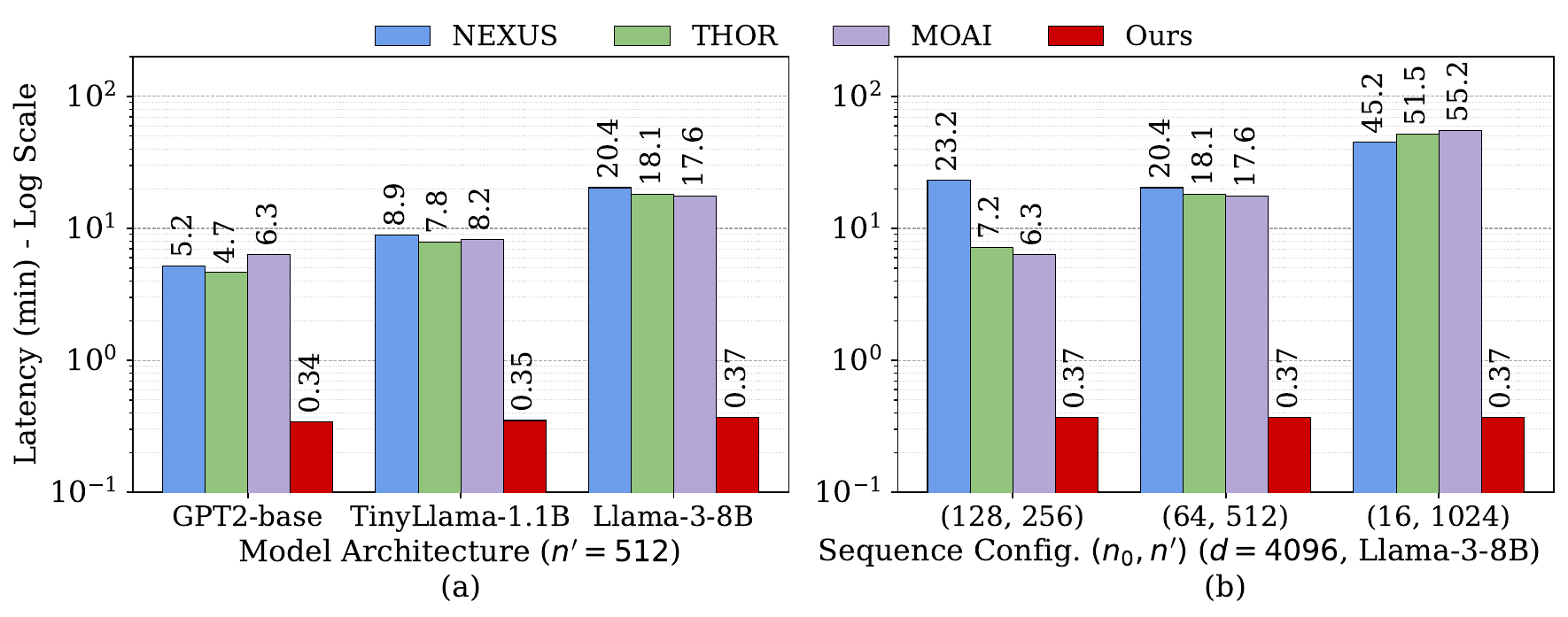}
  \caption{Latency comparison of a decoding step in one self-attention module measured in minutes. (a) Comparison across different model architectures with a fixed sequence length $n'=512$. (b) Comparison across different sequence length configurations $(n_0, n')$ with a fixed hidden dimension $d=4096$ (corresponding to Llama-3-8B).}
  \label{fig:kv_cache_performance}
\end{figure}

\begin{table}[!tb]
\centering
\caption{Models used in our evaluation.}
\label{tab:model_info}
\footnotesize
\begin{threeparttable}
    \centering
    \begin{tabular}{c|c|c|c}
    \toprule
    \textbf{Model} & \textbf{\#Heads} & \textbf{Hidden dim.} & \textbf{\#layers} \\ 
    \midrule
    GPT2-base \cite{radford2019languagegpt2} & 12 & 768 & 12 \\ 
    TinyLlama-1.1B \cite{zhang2024tinyllama} & 16 & 2048 & 22 \\ 
    Llama-3-8B \cite{grattafiori2024llama3herdmodels} & 32 & 4096 & 32 \\ \bottomrule
\end{tabular}
\end{threeparttable}
\end{table}

\subsection{Micro-benchmarks}

We first verify the effectiveness of our encoding methods, namely KV cache compatible encoding and efficient VMM algorithm. We set the input level of all modules to $4$ throughout the subsection to ensure a fair comparison.

\textbf{Complexity reduction with KV cache.}
Figure \ref{fig:kv_cache_performance} shows the latency of KV cache under different configurations, in which our method consistently outperforms the baselines by $14\sim149\times$ across different models and sequence lengths.

\begin{table}[!tb]
\centering
\caption{Comparison of VMMs with different hidden dimensions. We measure time in seconds.}
\label{tab:micro_ctpt}
    \footnotesize
  \begin{threeparttable}
    \begin{tabular}{cc|c|c|c}
    \toprule
    \multicolumn{2}{c}{Methods} & 768 & 2048 & 4096 \\
    \midrule
    \multirow{4}{*}{\textbf{Bolt\cite{pang2023bolt}+Padding}} & \#Rot & 375 & 375 & 375 \\
    & l=1 & 2.57 & 3.90 & 6.03 \\
    & l=4 & 6.13 & 9.85 & 15.80 \\
    & l=13 & 26.79 & 42.80 & 68.44 \\
    
    \midrule
    \multirow{4}{*}{\textbf{direct w/ BSGS}} & \#Rot & 108 & 188 & 252 \\
    & l=1 & 2.12 & 5.11 & 9.73 \\
    & l=4 & 5.62 & 13.78 & 26.46 \\
    & l=13 & 24.30 & 59.55 & 114.25 \\
    
    \midrule
    \multirow{4}{*}{\textbf{Cachemir VMM}} & \#Rot & 20 & 30 & 52 \\
    & l=1 & 0.12 & 0.28 & 0.61 \\
    & l=4 & 0.59 & 0.91 & 1.33\\
    & l=13 & 2.61 & 3.98 & 4.31\\
    \bottomrule
    \end{tabular}%
\end{threeparttable}
\end{table}
 

\begin{table*}[t]
\caption{Performance breakdown of a typical decoder in one generation step of LLama-3-8B \cite{grattafiori2024llama3herdmodels}. To verify the effectiveness of each of our methods, we apply our methods one by one. We show five columns of data: (a) MOAI's baseline method without KV Cache; (b) Thor's baseline method without KV Cache; (c) Cache only: naively applying KV Cache without any optimization; (d) Cache and VMM: Using both KV cache and our efficient VMM algorithm; (e) \method: Using all the methods including bootstrapping placement. We measure time in seconds, and assume the sequence length $(n_0, n') = (64, 512)$.}
\label{tab:main_exp}
  \footnotesize
  \begin{threeparttable}
    \begin{tabular}{c|ccccccccccc}
    \toprule
    \multirow{2}{*}{\textbf{Operation}}& 
    \multicolumn{2}{c}{MOAI\cite{zhang2025moai}} &
    \multicolumn{2}{c}{THOR\cite{moon2024thor}} &
    \multicolumn{2}{c}{Cache only} &
    \multicolumn{2}{c}{Cache \& VMM} & 
    \multicolumn{3}{c}{\method} \\
    &\textbf{Level} & \textbf{Time}  & \textbf{Level} & \textbf{Time} & \textbf{Level} & \textbf{Time} &  \textbf{Level} & \textbf{Time} & \textbf{Level} & \textbf{Time (CPU)} & \textbf{Time (GPU)}\\
    \midrule
    $\mathbf{Q}, \mathbf{K}, \mathbf{V}$ & 10$\rightarrow$13 & 1968.44 & 8$\rightarrow$13 & 2010.65 & 8$\rightarrow$13 & 68.89 & 10$\rightarrow$13 & 13.24 & 4 & 3.98 & 0.15 \\
    
    RoPE \& Cache & 12 & 2.61 & 11 & 2.68 & 11 & 8.22 & 12 & 8.29 & 3 & 2.71 & 0.01 \\
    
    $\mathbf{Q}\cdot \mathbf{K^T}$ & 11 & 693.27 & 10 & 329.85 & 10 & 4.44 & 11 & 4.91 & 2 & 0.79 & 0.09 \\
    
    Softmax & 10 & 158.32 & 8$\rightarrow$13 & 160.88 & 8$\rightarrow$13 & 20.11 & 9 & 19.78 & 0$\rightarrow$9 & 19.78 & 0.02 \\
    
    $\mathbf{Score}\cdot \mathbf{V}$ & 2 & 163.91 & 2 & 65.60 & 2 & 1.72 & 2 & 1.72 & 2 & 1.72 & 0.04 \\
    
    Output projection & 1 & 54.53 & 0$\rightarrow$13 & 670.22 & 0$\rightarrow$13 & 22.97 & 0$\rightarrow$13 & 4.42 & 0$\rightarrow$7 & 2.20 & 0.06 \\
    
    Add \& Norm & 0$\rightarrow$13 & 551.04 & 11 & 640.64 & 11 & 10.01 & 12 & 13.61 & 6 & 7.65 & 0.01 \\
    
    Up \& Gate projection & 6 & 2854.22 & 8 & 2777.67 & 8 & 90.64 & 9 & 9.98 & 3 & 2.96 & 0.34 \\
    
    SiLU & 5$\rightarrow$13 & 881.92 & 6$\rightarrow$13 & 3673.60 & 6$\rightarrow$13 & 14.35 & 8$\rightarrow$13 & 14.35 & 0$\rightarrow$13 & 14.35 & 0.01 \\
    
    Down projection & 6 & 1427.11 & 1$\rightarrow$13 & 2666.80 & 1$\rightarrow$13 & 87.26 & 1 & 0.73 & 1 & 0.73 & 0.14 \\
    
    Add \& Norm & 5$\rightarrow$13 & 551.04 & 11 & 640.64 & 11 & 10.01 & 0$\rightarrow$13 & 14.35 & 0$\rightarrow$7 & 8.66 & 0.01 \\
    
    Amortized Prefilling & / & 6.51 & / & 8.95 & / & 91.40 & / & 58.06 & / & 53.09 & 0.38 \\
    
    Bootstrappings & / & 14021.84 & / & 18439.68 & / & 384.21 & / & 360.20 & / & 360.20 & 1.78 \\
    
    Total & / & 23332.15 & / & 32087.85 & / & 822.58 & / & 522.57 & / & 477.81 & 3.02 \\
    \bottomrule
    \end{tabular}%
\end{threeparttable}
\end{table*}

\textbf{Effectiveness of VMM protocol.}
Table \ref{tab:micro_ctpt} presents the latency and operation number comparison between the prior implementation of VMM and our method. Our method substantially reduces the number of rotations and the latency by $7\sim19\times$ and $10\sim21 \times$.

\textbf{Acceleration with simplified bootstrapping placement.} We run the bootstrapping placement algorithm with our DAG simplification algorithm and with the baseline method of Orion\cite{Ebel2023OrionAF}, the latency being $234.24$ and $463.31$ milliseconds, respectively, achieving an $1.98\times$ acceleration.
We argue that the performance gain increases the ability of FHE compilers to scale to larger models, such as generative language ones.

\subsection{End-to-end Performance Evaluation}

\textbf{End-to-end performance breakdown.}
A typical end-to-end latency breakdown, as well as the ablation study, is shown in Table \ref{tab:main_exp}. For methods not using our bootstrapping placement algorithm, we use a straightforward method that only executes bootstrapping when necessary and bootstraps to the highest level by default.


Applying KV cache drastically accelerates the inference by $28.36\times$ and $39.01\times$ by reducing the complexity of all operations. However, a considerable portion of the gain is negated, with a gap of magnitudes of orders between the measured acceleration ratio and the theoretical ratio under plaintext. This is due to the significant bottleneck of ciphertext-plaintext multiplications, which makes up $77.75\%$ of the decoding latency except bootstrapping and is caused by slot underutilization. Therefore, further application of our efficient VMM algorithm mitigates both computational burden and level consumption of linear layers, resulting in an acceleration of $1.57\times$. Moreover, application of our tailored bootstrapping placement algorithm optimizes the global level assignment, creating an additional $1.10\times$ acceleration. Overall, \method~ provides a latency reduction of $48.83\times$ and $67.16\times$ compared to MOAI\cite{zhang2025moai} and THOR\cite{moon2024thor}, respectively.

When further accelerated by our GPU implementation, most homomorphic primitives are accelerated by $2\sim3$ orders of magnitudes. A detailed benchmark for these primitives are shown in \ref{apd:gpu}. This is completely orthogonal to our optimization and brings about an additional $158\times$ speedup. With all the optimizations above, generating one token for Llama-3-8B only consumes $1.61$ minutes, in comparison of Thor's\cite{moon2024thor} reported $10.43$ minutes for a BERT-base model. The results show the potential and scalability of our method for real-world deployment.

\begin{figure}[!t]
  \centering
  \includegraphics[width=\linewidth]{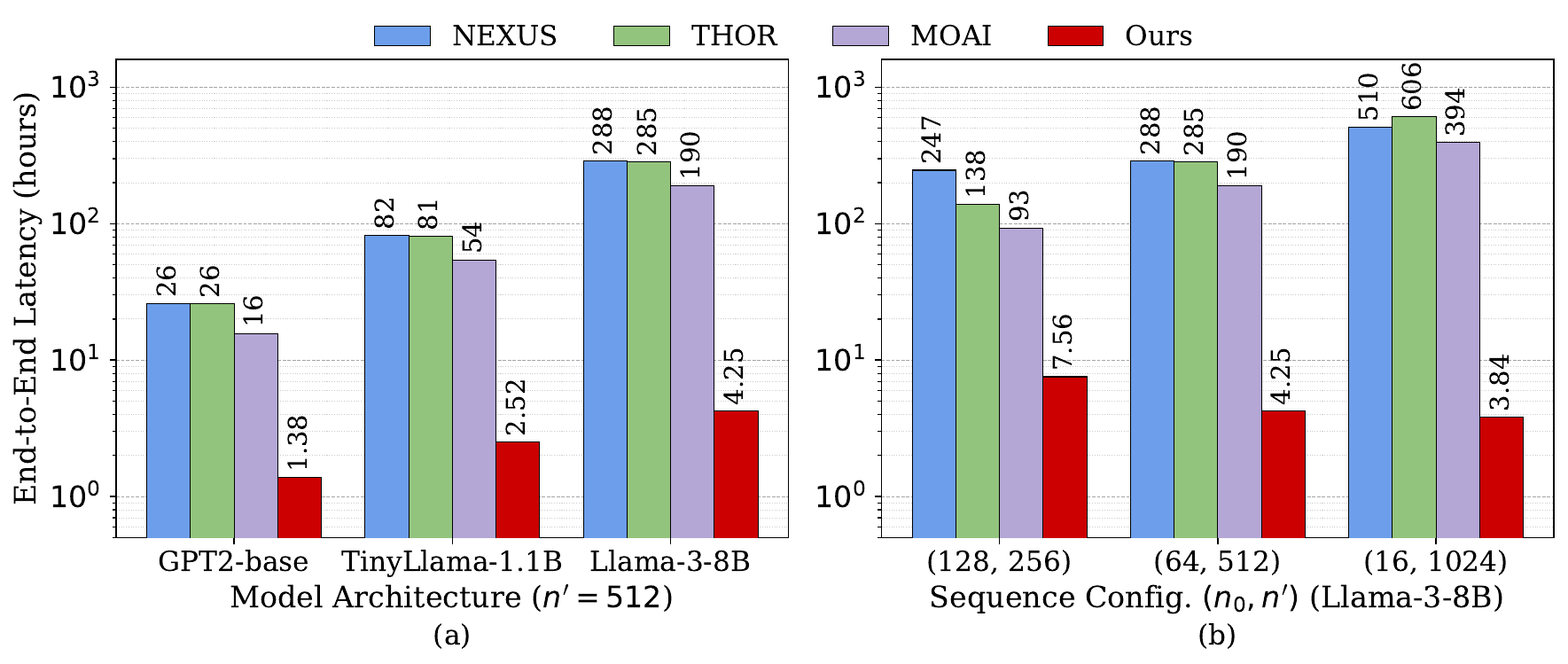}
  \caption{End-to-end latency comparison (measured in hours). (a) Latency across different model architectures with fixed sequence lengths $(n_0, n') = (64, 512)$. (b) Latency across different sequence configurations using Llama-3-8B. }
  \label{fig:e2e_latency}
\end{figure}

\textbf{End-to-end latency under different configurations.}
To further exemplify the efficacy of our method, we present the end-to-end latency evaluation with different models, input sequence lengths, and total sequence lengths. 
We present the end-to-end comparison between our methods and the prior-art baselines in Figure \ref{fig:e2e_latency}, where our method generates a significant latency reduction of $11\sim133\times$ for all cases. The key observation is that our method works better in larger models and longer sequences. The larger the hidden dimension is, the more dominant the savings is of our VMM algorithm; similarly, the more tokens there are, the more efficient it is to utilize the KV cache in both plaintext and ciphertext. Therefore, we argue that our method is more tailored to the real-world application scenario, where both large models and long contexts become mainstream.

\subsection{Discussions on Accuracy}

\begin{table}[!tb]
\centering
\caption{Accuracy of different FHE approximations of non-linear layers. The precision is computed as the negative logarithm of the root mean square error.}
\label{tab:module_acc}
  \footnotesize
  \begin{threeparttable}
    \begin{tabular}{@{\hspace{0.6em}}c@{\hspace{0.6em}}|@{\hspace{0.4em}}c@{\hspace{0.6em}}c|@{\hspace{0.4em}}c@{\hspace{0.6em}}c|@{\hspace{0.4em}}c@{\hspace{0.6em}}c@{\hspace{0.6em}}}
    \toprule
    \multirow{2}{*}{\textbf{Module}}  & \multicolumn{2}{c}{\textbf{NEXUS}} & \multicolumn{2}{c}{\textbf{THOR (Ours)}} & \multicolumn{2}{c}{\textbf{MOAI}} \\

     & depth & precision & depth & precision & depth & precision \\
    \midrule
    
    Softmax & 16 & 5.73 & 43 & 14.99 & 20 & -17.21 \\
    Norm & 16 & -31.85 & 16 & 10.81 & 20 & 9.69 \\
    SiLU & 14 & 2.40 & 10 & 27.54 & 7 & 26.78 \\
    \bottomrule
    \end{tabular}%
\end{threeparttable}
\end{table}

\begin{table}[!tb]
\centering
\caption{End-to-end performance of the approximated Llama-3-8B model.}
\label{tab:model_acc}
  \footnotesize
  \begin{threeparttable}
    \begin{tabular}{@{\hspace{0.6em}}c@{\hspace{0.6em}}|@{\hspace{0.4em}}c@{\hspace{0.6em}}c|c@{\hspace{0.6em}}c}
    \toprule
    \multirow{2}{*}{\textbf{Model}}  & \multicolumn{2}{c|}{\textbf{MRPC}} & \multicolumn{2}{c}{\textbf{Wikitext}}\\

     & accuracy & F1-score & byte ppl & word ppl \\
    \midrule
    
    plaintext & 71.32 & 82.19 & 1.50 & 8.83 \\
    approximated & 71.57 & 82.31 & 1.50 & 8.83 \\
    \bottomrule
    \end{tabular}%
\end{threeparttable}
\end{table}

We now briefly discuss on the accuracy of our methods.
For non-linear layers, \method~uses the scheme of THOR\cite{moon2024thor}, and a module-wise precision comparison is shown in Table~\ref{tab:module_acc}. Both schemes of NEXUS\cite{Zhang2024SecureTI} and MOAI for Softmax are problematic due to the usage of direct Taylor expansion for exponentiation, which is incompatible with the large input range in Llama-3-8B without scaling. Moreover, normalization of NEXUS also lacks a scaling factor after the summation along hidden dimension, as pointed out by MOAI. Despite the relatively high multiplicative depth, THOR\cite{moon2024thor} exhibits the most reasonable accuracy for accurate end-to-end inference.

We also include Table \ref{tab:model_acc} to show that our adapted non-linear approximation has enough precision for both classification and generative tasks, and for larger models like Llama-3-8B. As demonstrated, as long as the error of non-linear layers are bounded, the model performance is only negligibly affected or even improved in both datasets and both metrics without further downstream finetuning. This validates the practicality of our method.

\subsection{Memory Usage}

\begin{figure}[t]
  \centering
  \includegraphics[width=0.9\linewidth]{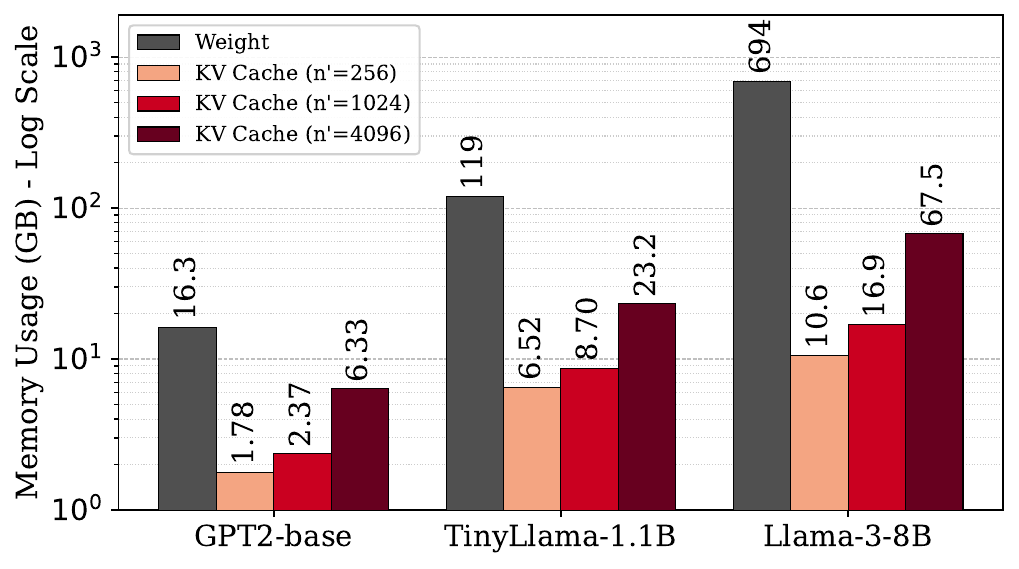}
  \caption{Memory footprint breakdown comparing static model weights and dynamic KV cache under varying sequence lengths ($n'$). }
  \label{fig:mem_usage}
\end{figure}



For Llama-3-8B, with our parameter setting, the KV cache of 1024 tokens consumes around 17 GB. In contrast, the weight plaintexts consume hundreds of GB. We show a more detailed comparison in Figure \ref{fig:mem_usage}. For larger models, the memory of weight plaintexts will become even more dominating as the encrypted KV cache size scales linearly with model dimensions while the weight plaintext size scales quadratically. When the sequence length becomes extremely large, the memory of encrypted KV cache becomes comparable to weight plaintexts. However, there are KV cache compression algorithms proposed in existing works \cite{liu2023scissorhandsexploitingpersistenceimportance,ge2024modeltellsdiscardadaptive}, which will become an interesting topic for future research. 





\section{Related Work}
\label{sec:rel}

\subsection{Private Transformer Inference}
\supp{Several studies have investigated how to enable private Transformer inference. In the MPC-based approach, Privformer \cite{akimoto2023privformer}, Puma \cite{dong2023puma}, SIGMA \cite{gupta2023sigma} present three-party protocols that require additional trust assumptions. 
There are works based on hybrid HE/MPC. Iron \cite{hao2022iron} first optimizes the packing for ct-pt matrix multiplication and non-linear layer protocols. Building on this, BOLT \cite{pang2023bolt} and Bumblebee \cite{lu2023bumblebee} further optimize the packing strategies and propose accurate linear approximation for non-linear layers. CipherGPT \cite{hou2023ciphergpt} optimizes matrix multiplication using VOLE-style OT \cite{Boyle2019EfficientTO} to reduce the amortized overhead of decoding, but it lacks support for KV Cache and performs a lot of redundant computations. Although these interactive approaches achieve faster execution times in the local area network setting, they still incur high communication costs due to the inherent overhead of MPC.}

\supp{FHE-based frameworks have also been explored for private Transformer inference. NEXUS \cite{Zhang2024SecureTI} is the first FHE-based framework that features novel HE-based computation of linear and non-linear operators, and achieves non-interactivity. However, NEXUS's amortizable offline-online computing strategy for ct-pt matrix multiplication suffers from high cost of preprocessing the weight matrix in decoding stage of inference. Among other works, PowerFormer \cite{park2024powerformer} improves upon NEXUS by leveraging the ct-ct matrix multiplication method from Jiang et al. \cite{jiang2018secure}. THOR \cite{moon2024thor} proposes a new diagonal packing algorithm to further reduce the HE rotations but at the cost of higher multiplicative depth. MOAI \cite{zhang2025moai} proposes interleaved batching packing to further reduce the number of rotations. However, these FHE-based works mainly focus on prefilling stage or single output tasks. Efficient KV-cache-enabled decoding inference for generative tasks is still a great challenge, as the efficient updating and usage of KV cache in attention module raise new demands for packing method. We address the problem by a dedicated KV-cache-aware packing method and an interleaved algorithm for efficient VMM computation. }

\subsection{Bootstrapping Placement}
Bootstrapping placement algorithms are mainly discussed in prior works on FHE compilers rather than FHE transformer inference, such as DACAPO\cite{compiler_dacapo}, Fhelipe\cite{compiler_fhelipe} and Orion\cite{Ebel2023OrionAF}. 
DACAPO\cite{compiler_dacapo} places bootstrappings through a two-stage scheme, first narrowing the search space down to a set of bootstrapping candidates and then comparing the latency for those different configurations, which introduces a high complexity and a prolonged runtime. Fhelipe\cite{compiler_fhelipe} uses dynamic programming to search for an efficient level assignment, completing the process in a complexity of $Ld\log d$. However, it naively omits all the shortcuts, introducing a high risk of missing the optimal solution. As described in detail in Section \ref{sec:bootstrap}, Orion\cite{Ebel2023OrionAF}, which models the problem into a shortest path search in a given DAG, boasts the SOTA algorithm. Still, it suffers from the aforementioned drawback that it restricts bootstrappings to the layer boundaries.
Our adapted method, by contrast, tackles the problems and provides a general solution for transformer models.

\section{Conclusion}
\label{sec:concl}

In this work, we propose \method, a KV cache accelerated homomorphic encrypted LLM inference regime for secure LLM inference. \method~features a set of novel HE packing algorithms to support KV-cache-enabled decoding, an interleaved replicated packing algorithm for efficient VMM computation, and an augmented bootstrapping placement algorithm. Extensive experimental results demonstrate that \method~ achieves significant speedup over prior-art frameworks for various models and sequence lengths with little impact on memory overhead or accuracy. We believe that \method~represents a pivotal step towards the real-world deployment of secure LLM inference.





\section{Ethical Considerations}

This work focuses on improving the efficiency of privacy-preserving inference frameworks. Our contributions aim to advance secure computation techniques without introducing new risks to data security or user privacy. By improving the practicality and scalability of privacy-preserving machine learning, we seek to foster the responsible and broader adoption of privacy-centric technologies.

We strictly adhere to the ACM Code of Ethics. Specifically, our work is motivated by the principles of respecting user privacy (Principle 1.6) and avoiding harm (Principle 1.2). All models and datasets used in our methodology are publicly available and do not involve proprietary, sensitive, or personally identifiable information. Our scheme does not introduce a new privacy risk with isolated KV Caches for different clients. We have carefully evaluated the ethical implications and ensure no privacy violations result from our experimentation or proposed methods.

To conclude, as the research involves no proprietary, sensitive, or personally identifiable information and does not involve human subjects, it is exempt from formal ethical review within our organization. By improving the practicality of privacy-preserving machine learning, we believe that this work can foster the responsible adoption of privacy-enhancing technologies, upholding data minimization and confidentiality. 

\bibliographystyle{ACM-Reference-Format}
\bibliography{references}

\cleardoublepage
\appendix
\setcounter{section}{0}
\section{Open Science}
To promote availability, all pre-trained models used in this work are publicly accessible through their sources, as referenced in the paper. 
Our source code and materials for replicating experiments are publicly available at \hyperlink{here}{https://anonymous.4open.science/r/Cachemir-B254}.
We welcome feedback and contributions to improve the implementation and further the research in privacy-preserving inference.

\section{Detailed Process of Direct Packing and Replicated Packing}
\label{apd:D_R_packing}
\subsection{Direct Packing}
We can directly utilize the diagonal method. The input vector $A$ is directly packed into a ciphertext (e.g., $[a_0,a_1,a_2,a_3,0,0,0,0]$) and multiply with one diagonal of the plaintext weight matrix $B$ at a time, as shown in Figure \ref{fig:gemv}(c).

To compute the partial sum corresponding to different diagonals of $B$, rotations are needed to generate other permutations of the input vector (e.g., $[a_1,a_2,a_3,a_0]$).
However, in the direct packing scenario, a standard rotation operation $\text{Rot}_l(1)$ would cause wrapped values like $[a_1,a_2,a_3,0,0,0,0,a_0]$. To avoid this, we can use \textbf{inner rotation} (Figure \ref{fig:gemv}(b)). A single inner rotation requires 2 rotations and 2 ct-pt multiplications, and consumes an extra multiplication depth. 
Overall, the entire computation using direct packing requires \textbf{$2d$ rotations} and \textbf{2 multiplication depths}, which is suboptimal due to wasted slots and the overhead of inner rotations.

\subsection{Replicated Packing}
To improve slot utilization of direct packing, a natural improvement is to replicate the input vector $A$ to fill all available slots. We term this approach \textbf{replicated packing}, shown in Figure \ref{fig:gemv}(d).
For example, the vector becomes $[a_0,a_1,a_2,a_3,a_0,a_1,a_2,a_3]$, enabling simultaneous multiplication with multiple diagonals of the matrix.

However, this method requires extra inner rotations to align the replicated vectors with their corresponding plaintext matrix diagonals.
For instance, to compute products with diagonals diag0 and diag1, $N/d-1$ inner rotation is performed to transform the ciphertext into a permuted form like $[a_0,a_1,a_2,a_3,a_1,a_2,a_3,a_0]$. This permuted vector is then multiplied by the packed diagonals of $B$. Then, similar to direct packing, inner rotations are also needed for other diagonals computation (e.g., diag2 and diag3). Finally, partial sums from different diagnals are accumulated via $\log(N/d)$ rotations to obtain the replicated form of the result $C$, $[c_0,c_1,c_2,c_3,c_0,c_1,c_2,c_3]$. This method requires \textbf{$O(2d^2/N+2N/d+\log (N/d))\sim O(2d^2/N+2N/d)$ rotations} and consumes \textbf{3 multiplication depths}. Although it is more efficient than direct packing in slot usage, reliance on inner rotations results in higher multiplicative depth overhead.

\section{Generalized Efficient Vector-Matrix Multiplication}
\label{apd:gemv}

\begin{algorithm}
\caption{Generalized version of our vector-matrix multiplication algorithm.}
\label{alg:gemv}
\begin{algorithmic}[1]
    \REQUIRE An interleaved activation ciphertext $\mathbf{c}^x$ that encrypts $\mathbf{x}\in\mathbb{R}^d\cup\mathbb{R}^{\alpha d}$, a vector of weight plaintext $\mathbf{p}_{j,k}$ where $\mathbf{W}\in\mathbb{R}^{d\times\alpha d}\cup\mathbb{R}^{\alpha d\times d}$. $\alpha, r_i\in\{ 2^n \mid n \in \mathbb{N} \}$. $r_o=\alpha/ri, t'=N/\alpha d, t=N/d$.
    \ENSURE An interleaved activation ciphertext $\mathbf{c}^y$ that encrypts $\mathbf{y}=\mathbf{x}\cdot\mathbf{W}$.
    \IF{$\mathbf{W}\in\mathbb{R}^{d\times\alpha d}$}
        \STATE $(t_1, t_2)\gets(t, t')$
    \ELSE
        \STATE $(t_1, t_2)\gets(t', t)$
    \ENDIF
    \FOR{$i=0$ to $\log t_1-1$}
        \STATE $\mathbf{c}'^x\gets\mathbf{c}^x\oplus \text{rot}(\mathbf{c}^x, 2^i(d-1))$
    \ENDFOR
    \FOR{$j=0$ to $r_i-1$}
        \STATE $\mathbf{c}_j''^x\gets\text{rot}(\mathbf{c}'^x, jt')$
        \STATE $\mathbf{c}'^y_k\gets\sum\limits_{j=0}^{r_i-1}\mathbf{c}''^x_j\otimes\mathbf{p}_{j,k}$
    \ENDFOR
    \STATE $r_o\gets d/r_i$
    \FOR{$k=1$ to $r_o$}
        \STATE $\mathbf{c}'^y_{r_o-k-1}\gets\mathbf{c}'^y_{r_o-k-1}\oplus\text{rot}(\mathbf{c}'^y_{r_o-k}, tr_i)$
    \STATE $\mathbf{c}^y\gets\mathbf{c}'^y_{0}$
    \ENDFOR
    \FOR{$l=0$ to $\log t_2 - 1$}
        \STATE $\mathbf{c}^y\gets\mathbf{c}^y\oplus\text{rot}(\mathbf{c}^y, 2^l)$
    \ENDFOR
    \RETURN $y$
\end{algorithmic}
\end{algorithm}

We present the generalized and detailed protocol of our efficient vector-matrix multiplication algorithm. When $\alpha=1$, the weight degenerates to a square matrix. When $\mathbf{W}\in\mathbb{R}^{kd\times d}$, the ciphertext and plaintext should be encoded in the form where
$\mathbf{c}^x[i]=\mathbf{x}[\lfloor i/t\rfloor+(i/t')\%\alpha\cdot d]\mathbb{I}\{t'\mid i\}$
and 
$\mathbf{p}_{j,k}[i]=\mathbf{W}[(\lfloor i/t\rfloor+(i/t')\%\alpha\cdot d+i\%t+jt)\%(\alpha d), (i/t-ktr_i)\%d]$
, respectively. Similarly, when $\mathbf{W}\in\mathbb{R}^{d\times kd}$, we need $\mathbf{c}^x[i]=\mathbf{x}[i/t]\mathbb{I}\{t\mid i\}$ and $\mathbf{p}_{j,k}[i]=\mathbf{W}[(i/t+i\%t+jt)\%d, (\lfloor i/t\rfloor+(i/t')\%\alpha\cdot d-ktr_i)\%(\alpha d)]$. We permute the output of up projection to enable an interleaved reduction during down projection, so that the encoding format is kept the same throughout the linear layers.

Throughout the algorithm, we execute $\log t+\log t'+r_i+r_o$ times of rotations, which gives the complexity of $O(\log(\alpha N^2/d^2)+2\sqrt{\alpha d^2/N}$ because $r_i\cdot r_o\cdot t'=d$. Notably, more than half of the rotations are hoisted. 

\section{Implementation of RoPE}
\label{apd:RoPE}
Now we brief on our implementation of RoPE. We first encoder three plaintexts with different trigonometric functions:
\begin{align}
\mathbf{p}_0^{\text{RoPE}}[i]&=\cos(\theta(i, n))\mathbb{I}\{t\mid i\}\\
\mathbf{p}_1^{\text{RoPE}}[i]&=\sin(\theta(i, n))\mathbb{I}\{2t\mid i\}\\
\mathbf{p}_2^{\text{RoPE}}[i]&=-\sin(\theta(i, n))\mathbb{I}\{2t\mid (i+t)\}
\end{align}

Then, we can evaluate RoPE with the following equation:
\[
\mathbf{c}^y=\mathbf{c}^x\otimes\mathbf{p}_0^{\text{RoPE}}+\text{rot}(\mathbf{c}^x\otimes\mathbf{p}_1^{\text{RoPE}}, 1)+\text{rot}(\mathbf{c}^x\otimes\mathbf{p}_2^{\text{RoPE}}, -1)
\]
which simultaneously extracts useful slots with only 3 multiplications, 2 rotations, and 1 multiplicative depth.

\section{Non-linear Function Implementation}
\label{apd:nonlinear}

We largely follow the non-linear function implementations in the open-source THOR~\cite{moon2024thor}. However, since our evaluation targets larger models than those considered in THOR, its original parameter configurations are no longer directly applicable. To this end, we first profile the runtime input ranges of both Softmax, LayerNorm and SiLU, and use these empirical statistics to guide our design choices. Part of the concrete input ranges are listed in Table \ref{tab:input_range}.

Concretely, the profiled input ranges inform the selection of the initial exponential approximation of the form $c \cdot \exp(kx)$ used by approximate Softmax~\cite{cho2024fast}, as well as the scaling factors applied to LayerNorm inputs. These scaling steps normalize operands prior to approximate inversion, ensuring that all inputs lie within the interval $(0,1)$. We deliberately set the upper bound to be close to, but not exceeding, $1$, which in turn prevents the effective lower bound from becoming excessively small. Such normalization reduces the multiplicative cost of inversion and improves numerical accuracy.

While our overall design largely aligns with THOR, we diverge in the choice of inversion strategy. THOR employs an aSOR-based method~\cite{moon2024adaptive}, whereas we compute approximate reciprocals using the standard Goldschmidt algorithm. Although aSOR can reduce level consumption by explicitly managing scales and avoiding additional multiplications from adaptive correction factors, it introduces increased uncertainty in level usage. In contrast, the standard Goldschmidt method offers a predictable and profiled iteration depth, which better aligns with our proposed bootstrapping placement strategy (Section~\ref{sec:bootstrap}), while still providing sufficient numerical accuracy for all evaluated workloads.

\section{GPU Implementation of RNS-CKKS}
\label{apd:gpu}
To provide high-performance cryptographic primitives for Cachemir, we develop our GPU-accelerated RNS-CKKS implementation by customizing the open-source Phantom library~\cite{yang2024phantom} with several advanced GPU optimization techniques. Table~\ref{tab:ckks_microbench} reports the microbenchmarks for fundamental homomorphic operations and bootstrapping under the parameter configurations defined in Section~\ref{ssec:eval-step}.
\begin{table}[!t]
\centering
\caption{Input ranges of non-linear layers.}

\label{tab:input_range}
\footnotesize
\begin{threeparttable}
    \begin{tabular}{c|c|cccc}
    \toprule
    \textbf{Module} & \textbf{Index} & \textbf{Mean} & \textbf{Std} & \textbf{Max} & \textbf{Min} \\
    \midrule
    \multirow{3}{*}{Softmax} & 1 & 1.78 & 3.54 & 23.12 & -24.38 \\
    & 6 & -3.18 & 2.57 & 10.5 & -18.5 \\
    & 31 & -3.04 & 5.75 & 67.50 & -17.63 \\
    \midrule
    \multirow{3}{*}{Norm} & 1 & 0.00 & 0.01 & 0.37 & -0.39 \\
    & 6 & 0.00 & 0.99 & 8.31 & -31.75 \\
    & 31 & 0.00 & 1.24 & 15.81 & -48.25 \\
    \midrule
    \multirow{3}{*}{SiLU} & 1 & -0.04 & 0.10 & 3.97 & -2.94 \\
    & 6 & -0.26 & 0.32 & 6.13 & -7.22 \\
    & 31 & -0.02 & 1.28 & 24.50 & -31.63 \\
    \bottomrule
    \end{tabular}
\end{threeparttable}
\end{table}

\begin{table}[!t]
\centering
\caption{Performance microbenchmarks of RNS-CKKS primitive operations and bootstrapping. For primitives, $L$ is the current level (us); for bootstrapping, $L$ is the target level (ms). Parameters are detailed in Section~\ref{ssec:eval-step}.}
\label{tab:ckks_microbench}
\footnotesize
\begin{threeparttable}
    \begin{tabular}{lrrrrr}
    \toprule
    \textbf{Operation} & \textbf{$L=1$} & \textbf{$L=4$} & \textbf{$L=7$} & \textbf{$L=10$} & \textbf{$L=13$} \\
    \midrule
    ct-ct Addition (us)   & 9.51   & 11.86  & 13.10  & 27.87  & 35.27  \\
    ct-pt Multiplication (us)  & 66.07  & 87.76  & 158.22 & 140.82 & 178.73  \\
    ct-ct Multiplication (us) & 223.23 & 319.28 & 459.76 & 584.87 & 816.87 \\
    Rotation (us)      & 164.35 & 234.91 & 301.46 & 481.62 & 608.48 \\
    \midrule
    Bootstrapping (ms)  & 51.30  & 66.88  & 85.39  & 104.92 & 125.58 \\
    \bottomrule
    \end{tabular}
\end{threeparttable}
\end{table}

\end{document}